\documentclass[aps,nofootinbib,superscriptaddress,notitlepage,preprintnumbers,10pt]{revtex4-1}
\pdfoutput=1
\usepackage{graphicx}
\usepackage{amsmath,amssymb}
\usepackage{hyperref}
\usepackage[usenames]{color}
\usepackage{url}
\usepackage{epstopdf}
\allowdisplaybreaks	
\numberwithin{equation}{section}

\hypersetup{
    colorlinks=true,
    linkcolor=black,
    citecolor=blue,
    urlcolor=black
}

\def\be{\begin{equation}}
\def\ee{\end{equation}}
\def\ba{\begin{eqnarray}}
\def\ea{\end{eqnarray}}
\def\nn{\nonumber}
\def\f{\frac}
\def\l{\left}
\def\r{\right}

\begin{document}
\title{On the stability conditions for theories of modified gravity \\ in the presence of matter fields}
\author{Antonio De Felice}
\thanks{E-mail:antonio.defelice\textit{@}yukawa.kyoto-u.ac.jp}
\affiliation{Center for Gravitational Physics, Yukawa Institute for Theoretical Physics, Kyoto University, 606-8502, Kyoto, Japan}
\affiliation{The Institute for Fundamental Study ``The Tah Poe Academia Institute'',
Naresuan University, Phitsanulok 65000, Thailand}
\author{Noemi Frusciante}\thanks{E-mail:fruscian\textit{@}iap.fr}
\smallskip
\affiliation{$^{}$ Sorbonne Universit$\acute{\text{e}}$s, UPMC Univ Paris 6 et CNRS, UMR 7095, Institut d'Astrophysique de Paris, GReCO, 98 bis bd Arago, 75014 Paris, France}
\smallskip 
\author{Georgios Papadomanolakis}\thanks{E-mail:papadomanolakis\textit{@}lorentz.leidenuniv.nl}
\affiliation{
$^{}$ Institute Lorentz, Leiden University, PO Box 9506, Leiden 2300 RA, The Netherlands}
\smallskip
\preprint{YITP-16-102}

\begin{abstract}
  \vspace{0.3cm}
  We present a thorough stability analysis of  modified gravity theories in the presence of matter fields. We use the  Effective Field Theory framework for Dark Energy and Modified Gravity to retain a general approach for the gravity sector and a Sorkin-Schutz action for the matter one. Then, we work out the proper viability conditions to guarantee in the scalar sector the absence of  ghosts, gradient and tachyonic instabilities. The absence of ghosts can be achieved by demanding a positive  kinetic matrix, while the lack of a gradient instability is ensured by imposing a positive speed of propagation for all the scalar modes. In case of tachyonic instability, the mass eigenvalues have been studied and we work out the appropriate expressions. For the latter, an instability occurs only when the negative mass eigenvalue is much larger, in absolute value, than the Hubble parameter. We discuss the results for the minimally coupled quintessence model showing for a particular set of parameters two typical behaviours which in turn lead to a stable and an unstable configuration. Moreover, we find that the speeds of propagation of the scalar modes strongly depend on matter densities, 
  for the beyond Horndeski theories. 
 Our findings can be directly employed when testing modified gravity theories as they allow to identify the correct viability space.
\end{abstract}
\vspace{0.2cm}
\date{\today}
\maketitle
\flushbottom

\tableofcontents

\section{Introduction}

Since its discovery, the late-time cosmic acceleration phenomenon has been the most challenging problem for cosmologists. Usually referred to  as the Dark Energy (DE) problem, at first it was explained with the presence of a cosmological constant ($\Lambda$) in  General Relativity (GR).
The resulting standard cosmological model ($\Lambda$CDM) offers an exquisite fit to cosmological data~\cite{Ade:2015xua}, however it suffers from some  major theoretical issues which are still unresolved (see ref.~\cite{Bull:2015stt} and references therein).  This has paved the way to new theories of gravity to be considered as valid alternative to GR~\cite{Sotiriou:2008rp,Silvestri:2009hh,DeFelice:2010aj,Clifton:2011jh,Tsujikawa:2013fta,Deffayet:2013lga,Joyce:2014kja,Koyama:2015vza,Bull:2015stt}. Such theories include mechanisms able to give rise to the observed acceleration at large scales and late time, while being hidden at solar system scale~\cite{Joyce:2014kja} where GR is well tested. A common aspect present in most of these modified gravity (MG) theories is to revise GR by including an additional scalar degree of freedom (DoF), whose dynamics can explain the current observations.

Irrespectively of the resulting MG model,  one has to ensure that the evolution of the modes associated to the extra DoF does not lead to pathological instabilities, such as ghost, gradient and tachyonic instabilities (for a review see ref.~\cite{Sbisa:2014pzo}).  In particular, when studying cosmological perturbations the additional DoF is coupled to one or
more DoFs representing the matter fields dynamics, then these couplings imply that a consistent and complete study of the stability of the whole system can not be done without considering the interaction with the matter sector.  In fact, the stability conditions might be altered by the presence of the additional matter fields, thus changing the
viability space of the theory~\cite{Scherrer:2004au,Bertacca:2007ux,Bertacca:2007cv,Gergely:2014rna,Kase:2014cwa,Gleyzes:2014qga}. Identifying the correct viability  requirements is important when testing MG theories with cosmological data by using statistical tools~\cite{Zhao:2008bn,Hu:2013twa,Raveri:2014cka,Zumalacarregui:2016pph}, as they can reduce the viability space one needs to explore. Additionally they can even dominate over the constraining power of observational data as recently shown in the case of designer f(R)-theory on $w$CDM background~\cite{Raveri:2014cka}. 

With the aim to obtain general results, we will employ the Effective Field Theory of Dark Energy and Modified Gravity (hereafter EFT) presented in refs.~\cite{Gubitosi:2012hu,Bloomfield:2012ff}. Inspired by the EFT of Inflation~\cite{Creminelli:2006xe,Cheung:2007st,Weinberg:2008hq,Creminelli:2008wc} and large scale structures~\cite{Park:2010cw,Jimenez:2011nn,Carrasco:2012cv,Hertzberg:2012qn,Carrasco:2013mua,Porto:2013qua,Senatore:2014vja}, it was  widely studied  in refs.~\cite{Gleyzes:2013ooa,Bloomfield:2013efa,Piazza:2013coa,Frusciante:2013zop,Gleyzes:2014rba,Perenon:2015sla,Kase:2014cwa,Frusciante:2016xoj}. The EFT approach provides a model independent framework  to study linear order cosmological perturbations in theories of gravity which exhibit an additional scalar DoF, while at the same time it parametrizes in an efficient way existing models, since most of them can be directly mapped into this language~\cite{Gubitosi:2012hu,Bloomfield:2012ff,Gleyzes:2013ooa,Kase:2014cwa,Frusciante:2016xoj}.  Subsequently, the EFT approach has been implemented into the Einstein Boltzmann solver, CAMB/CosmoMC~\cite{CAMB,Lewis:1999bs,Lewis:2002ah}, creating EFTCAMB/EFTCosmoMC~\cite{Hu:2013twa,Raveri:2014cka,Hu:2014sea,Hu:2014oga,Frusciante:2015maa,Hu:2016zrh}  (\url{http://www.eftcamb.org/}),  providing a perfect tool  to test  gravity models through comparison with observational data. EFTCAMB comes with a built-in module to explore the viability space of the underlying theory of gravity, which then can be used as \textit{priors}.  The results of the present work have a direct application as they can be employed to improve the current EFTCAMB viability requirements but not limited to it as they can be easily mapped to other parametrizations~\cite{Hu:2014oga}.

The matter sector is  described by the Sorkin-Schutz action, which allows to treat general matter fluids~\cite{Sorkin:1977,Brown:1993}.
Among many models used to describe matter Lagrangians and which have been extensively used and investigated in the past years~\cite{Scherrer:2004au,Bertacca:2007ux,Bertacca:2007cv,Gergely:2014rna,Kase:2014cwa,Gleyzes:2014qga}, we choose to follow the recent arguments in ref.~\cite{DeFelice:2015moy}. Indeed, it has been shown that such an action, along with an appropriate choice for the matter field,  describes the dynamics of all matter 
fluids avoiding some problems which might arise when including pressure-less matter fluids, like dust or cold dark matter (CDM), which instead need to be considered as they are relevant in the evolution of the Universe. 

Recently, a stability analysis has appeared in the context of EFT~\cite{Kase:2014cwa}. However, in  our work we present also the conditions which allow to avoid tachyonic instabilities and we analyse, more in detail, in addition to the generic theories, all possible sub-cases concerning the stability conditions.  Furthermore, another difference comes with the choice of the matter Lagrangian, indeed in our analysis a pressure-less fluid can be safely considered.

With this machinery, we proceed to derive the viability constraints one needs to impose on the free parameters of the theory by focusing on three sources of possible instabilities,  ghost, gradient and tachyonic instabilities. We will proceed while retaining  the full generality of the EFT approach, i.e.\ without limiting to specific models. However, where relevant, we will make connections to specific theories, such as low-energy Ho\v rava gravity~\cite{Horava:2008ih,Horava:2009uw,Mukohyama:2010xz} and beyond Horndeski models~\cite{Gleyzes:2014dya} and we will analyse the results within the context of these models.
  
The present manuscript is organized as follows. In section~\ref{Sec:EFT}, we briefly recap the EFT formalism we use to parametrize the DE/MG models with one extra scalar DoF.  In section~\ref{Sec:matter}, we introduce the Sorkin-Schutz action to describe the dynamics of matter fluids and we discuss the advantage of using this action with respect to previous approaches. We also work out the corresponding continuity equation and  second order perturbed action. In section~\ref{Sec:stability}, we work out the action for both gravity and matter fields up to second order in perturbations. Then, we calculate and discuss the stability requirements to avoid ghost instabilities (section~\ref{Sec:ghost}), to guarantee  positive speeds of propagation (section~\ref{Sec:speed}) and to prevent tachyonic instabilities (section~\ref{Sec:Mass}). Finally, we conclude in section~\ref{Sec:conclusion}.

\section{The Effective Field Theory approach to Dark Energy and Modified Gravity}\label{Sec:EFT}

The  EFT of DE/MG  has been proposed as a  unifying framework to study the dynamic and evolution of linear order perturbations of a broad class of DE/MG theories~\cite{Gubitosi:2012hu,Bloomfield:2012ff}.  Indeed, this approach encloses all the theories of gravity  exhibiting one extra scalar and dynamical DoF  and  admitting a well-defined Jordan frame. The building blocks to construct the EFT action are the unitary gauge and the perturbations,  around the  Friedmann-Lema$\hat{\text{i}}$tre-Robertson-Walker (FLRW) background and up to second order, of all operators that are invariant under the time dependent spatial-diffeomorphisms. In front of each operator there is a time dependent function usually dubbed \textit{EFT function}. The explicit form of the perturbed EFT action is the following: 
\begin{align}\label{EFTaction}
\mathcal{S}_{EFT}&=\int d^4x\sqrt{-g}\l[\frac{m_0^2}{2}(1+\Omega(t))R^{(4)}+\Lambda(t)-c(t)\delta g^{00}+\frac{M^4_2(t)}{2}(\delta g^{00})^2-\frac{\bar{M}^3_1(t)}{2}\delta g^{00}\delta K-\frac{\bar{M}^2_2(t)}{2}(\delta K)^2\nonumber\r.\\
&\l.-\frac{\bar{M}_3^2(t)}{2}\delta K^{\mu}_{\nu}\delta K^{\nu}_{\mu}+\f{\hat{M}^2(t)}{2}\delta g^{00}\delta R^{(3)}+m^2_2(t)\l(g^{\mu\nu}+n^\mu n^\nu\r)\partial_{\mu}g^{00}\partial_{\nu}g^{00}\r],
\end{align}
where $m_0^2$ is the Planck mass, $g_{\mu\nu}$ is the four dimensional metric and  $g$  is its  determinant, $\delta g^{00}$ is the perturbation of the upper time-time component of the metric, $n_{\mu}$ is the normal vector to the constant-time hypersurfaces,  $R^{(4)}$ and $R^{(3)}$ are respectively the trace of the four dimensional and three dimensional Ricci scalar,  $K_{\mu\nu}$ is the extrinsic curvature and $K$ is its trace. Finally, with $\delta A= A-A^{(0)}$ we indicate the linear perturbation of the quantity $A$ and $A^{(0)}$ is the corresponding background value. 

The choice of the unitary gauge in the above action guarantees that the scalar DoF has been absorbed by the metric, hence it does not appear explicitly in the action. One could make it  manifest by applying the so called "St$\ddot{\text{u}}$ckelberg technique", thus disentangling the dynamics of the extra DoF from the metric one~\cite{Gubitosi:2012hu,Bloomfield:2012ff}. 

Moreover, it has been shown that  appropriate combinations of the EFT functions in action~(\ref{EFTaction}) allows one to describe specific classes of DE/MG models. We  group such combinations as follows:  
\begin{itemize}
\item ${M}_2^2=-\bar{M}_3^2=2\hat{M}^2$  and  $m_2^2=0$:  Horndeski~\cite{Horndeski:1974wa} or Generalized Galileon class of models~\cite{Deffayet:2009mn} (and all the models belonging to them);
\item  ${M}_2^2+\bar{M}_3^2=0$ and $m_2^2=0$ :     Beyond Horndeski class of models~\cite{Gleyzes:2014dya};
\item     $m_2^2\neq 0$:  Lorentz violating theories (e.g. low-energy Ho\v rava gravity~\cite{Horava:2008ih,Horava:2009uw,Mukohyama:2010xz}).
\end{itemize} 
For a detailed guide to map a specific theory into the EFT language we refer the reader to refs.~\cite{Gubitosi:2012hu,Bloomfield:2012ff,Gleyzes:2013ooa,Kase:2014cwa,Frusciante:2016xoj}. Finally, an extended version of the above EFT action has been presented in ref.~\cite{Frusciante:2016xoj} which includes  operators with higher than second order spatial derivatives.

In the following we will briefly recap the construction of the EFT  action up to second order  in terms of the scalar metric perturbations as it will be the starting point for the stability analysis.  For a  comprehensive review of the formalism  we refer the reader to the following papers~\cite{Kase:2014cwa,Frusciante:2016xoj}. 

Because of the  unitary gauge in action~(\ref{EFTaction}), it is natural to choose the Arnowitt-Deser-Misner (ADM) formalism~\cite{Gourgoulhon:2007ue} to write the line element, which reads: 
\begin{equation}
ds^2=-N^2 dt^2+h_{ij}(dx^i+N^i dt)(dx^j+N^j dt)\,,
\end{equation}
where $N(t,x^i)$ is the lapse function, $N^i(t,x^i)$ the shift and $h_{ij}(t,x^i)$ is the metric tensor of the three dimensional spatial slices.  Proceeding with the expansion around a flat FLRW background, the metric can be written as:
\be\label{scalarMetric}
ds^2=-(1+2\delta N)dt^2+2\partial_i\psi dtdx^i+a^2(1+2\zeta)\delta_{ij}dx^idx^j \,,
\ee
where as usual $\delta N(t,x^i)$ is the perturbation of the lapse function, $\partial_i\psi(t,x^i)$ and $\zeta(t,x^i)$ are the scalar perturbations respectively of $N_i$ and $h_{ij}$  and $a$ is the scale factor.  Then, the scalar perturbations of the quantities involved in the  action~(\ref{EFTaction}) are: 
\ba\label{scalarperturbations}
&&\delta g^{00}= 2\delta N \,,\nn\\
&&\delta K=-3\dot{\zeta}+3H\delta N+\f{1}{a^2}\partial^2\psi\,, \nn\\
&&\delta K_{ij}=a^2\delta_{ij}(H\delta N-2H\zeta-\dot{\zeta})+\partial_i\partial_j\psi\,, \nn\\
&&\delta K^i_j=(H\delta N-\dot{\zeta})\delta^i_j+\f{1}{a^2}\partial^i\partial_j\psi\,, \nn\\
&&\delta R^{(3)}=-\f{4}{a^2}\partial^2\zeta\,,
\ea
where we have made use of the following definitions of the normal vector and extrinsic curvature:
\begin{equation}\label{convention}
n_{\mu}=N\delta_{\mu }^0, \qquad K_{\mu\nu}=h^\lambda_\mu\nabla_{\lambda}n_{\nu},
\end{equation}
with $h^{\mu\nu}=g^{\mu\nu}+n^\mu n^\nu$, $H\equiv \f{1}{a}\f{d{a}}{dt}$ is the Hubble function and  dots are the derivatives with respect to time.
Then, the action~(\ref{EFTaction}) can be explicitly expanded in terms of metric scalar perturbations up to second order and after some manipulations, we obtain the following final form:
\begin{eqnarray}\label{EFT2}
  S_{EFT}^{(2)}=\int dt d^3x a^3
&\Biggl\{&
-{\frac {F_4 (\partial^2\psi)^{2}}{2{a}^{4}}}
-\frac32\,F_1\dot\zeta^2+m_0^2(\Omega+1)\,\frac{(\partial\zeta)^2}{a^2}-\frac{\partial^2\psi}{a^2}\left(F_2\delta N-F_1\dot\zeta\right)\nonumber\\
&&{}+4m_2^2\frac{[\partial(\delta N)]^2}{a^2}+\frac{F_3}2\,\delta N^2
+\left[3F_2\dot\zeta-2\l(m_0^2(\Omega+1)+2\hat{M}^2\r)\frac{\partial^2\zeta}{a^2}\right]\delta N\Biggr\},
\end{eqnarray}
where we have defined
\begin{eqnarray}
F_{1} & = & 2m_{0}^{2}(\Omega+1)+3\bar{M}_{2}^{2}+\bar{M}_{3}^{2}\,,\nn\\
F_{2} & = & HF_{1}+m_{0}^{2}\dot{\Omega}+\bar{M}_{1}^{3}\,,\nn\\
F_{3} & = & 4M_{2}^{4}+2c-3H^{2}F_{1}-6m_{0}^{2}H\dot{\Omega}-6H\bar{M}_{1}^{3}\,,\nn\\
F_4 &=& \bar{M}_2^2+\bar{M}_3^2\,, 
\end{eqnarray}
and other terms vanish because of the background equations of motion. This result will be  considered along with the matter sector which will be presented in the next section in order to facilitate the complete study of  conditions that guarantee that a gravity theory, in presence of matter fields, is free from instabilities.

\section{The Matter Sector}\label{Sec:matter}

The goal of the present work is to investigate the emergence of instabilities in modified theories of gravity under the influence of matter fluids and subsequently set appropriate stability conditions.  Therefore, a crucial step is to make the appropriate choice for the matter action, $S_m$. Moreover, the generality of the EFT approach in describing the gravity sector makes that even for the matter action there is an equally general treatment. It is common in literature to choose for the matter Lagrangian a k-essence like form, $P(\mathcal{X})$ \cite{Scherrer:2004au,Bertacca:2007ux,Bertacca:2007cv,Kase:2014cwa,DeFelice:2011bh,Gleyzes:2015pma,D'Amico:2016ltd}, to model the matter DoF where $\mathcal{X}\equiv\chi_{;\mu}\chi^{;\mu}$ is the kinetic term of the field $\chi$. However, this choice  displays problematic behaviours which motivates us to decide for a different action.
The easiest way of identifying those issues is to consider the corresponding action for $P(\mathcal{X})$ when it has been specialized to a dust fluid.  In that case it can  be easily shown that the action diverges. Subsequently, in ref.~\cite{DeFelice:2015moy}, it has been shown that the real problem arising in the K-essence like matter Lagrangian lies in the choice of the canonical field one uses to describe the DoFs of the fluid. Indeed,  the usual choice, as fluid variable, the velocity $v_m$, satisfies a first order, closed, equation of motion, which only requires one independent initial condition. Then, the dust fluid would have  only one DoF (rather than two) and for that field  the action tends to blow up as the speed of propagation goes to zero ($c_{s,d}^2 \rightarrow 0$).  Instead, the appropriate variable for the fluid is the matter density perturbation, $\delta_m$.
	 
In order to avoid the issues described above we choose the  Sorkin-Schutz action, see refs.~\cite{Sorkin:1977, Brown:1993} which is well defined for a dust component and can describe in full generality perfect fluids. As observed above the appropriate fluid variable is the density perturbation which is exactly the one employed by this action and thus satisfies a second order equation of motion as will be evident in the following. The  Sorkin-Schutz matter action reads:
\begin{equation}\label{matteraction}
S_{\mathrm{m}}=-\int d^{4}x[\sqrt{-g}\,\rho(n)+J^\nu\partial_\nu\ell]\,,
\end{equation}
where $\rho$ is the energy density, which depends on the number density $n$, $\ell$ is a scalar field,  whereas $J^\nu$ is a vector with weight one. Additionally, we define $n$ as
\begin{equation}\label{numberdensity}
n=\sqrt{\frac{J^\alpha J^\beta g_{\alpha\beta}}{g}}\,.
\end{equation}
Then, the four velocity vector $u^\alpha$ is defined as
\begin{equation}
u^\alpha=\frac{J^\alpha}{n\sqrt{-g}}\,,
\end{equation}
and satisfies the usual relation $u^\alpha u_\alpha=-1$. Variation of the matter Lagrangian with respect to $J^\alpha$ leads to
\begin{equation}
u_\alpha=\frac{1}{\partial\rho/\partial n}\,\partial_\alpha\ell\,,
\end{equation}
while taking its variation with respect to the metric we find that the stress energy tensor can be defined as
\begin{equation}
T_{\alpha\beta}\equiv \f{2}{\sqrt{-g}}\f{\delta S_m}{\delta g^{\alpha \beta}}  =n\,\frac{\partial\rho}{\partial n}\,u_\alpha u_\beta+\left( n\,\frac{\partial\rho}{\partial n}-\rho \right) g_{\alpha\beta}\,,
\end{equation}
which is a barotropic perfect fluid with pressure given by
\begin{equation}
p\equiv n\,\frac{\partial\rho}{\partial n}-\rho\,.
\end{equation}
Let us notice that a particular choice for the density, i.e.\ $\rho\propto n^{1+w}$, allows to have the usual relation $p=w\rho$, where $w$ is the barotropic coefficient.
Finally, by varying the matter action with respect to $\ell$, one gets the conservation constraint
\be
\partial_\alpha J^\alpha=0.  
\ee
On a flat FLRW background the above relation gives $J^0=\mathcal{N}_0$, where $\mathcal{N}_0$ is the total particle number and from Eq.~(\ref{numberdensity}) we have $n=\mathcal{N}_0/a^3$.

Let us now proceed to write the matter action~(\ref{matteraction}) up to second order in the scalar fields by using the metric scalar perturbations in Eq.~\eqref{scalarMetric}. For the fluid variables we proceed to expand them as follows
\begin{eqnarray}
J^0&=&\mathcal{N}_0+\delta J\,,\nn\\
J^i&=&\frac1{a^2}\,\partial^i\delta j\,,\nn\\
\ell&=&-\int^t \frac{\partial\rho}{\partial n}\,dt'-\frac{\partial\rho}{\partial n}\,v_m\,,
\end{eqnarray}
where $v_m$ is the velocity of the matter species. Furthermore, we note that since
\be
\rho=\bar{\rho}+\f{\partial\rho}{\partial n}\l(-\f{3\mathcal{N}_0}{a^3}\zeta+\f{\delta J}{a^3}\r)\equiv \bar{\rho}+\delta \rho \,,
\ee
where $\bar{\rho}$ is the density at the background, one can obtain
\begin{equation}
\delta J=\frac{a^3\bar{\rho}\,\delta_m}{\partial\rho/\partial n}+3\mathcal{N}_0\,\zeta\,,
\end{equation}
where, as usual, $\delta_m=\delta\rho/\bar{\rho}$. We can thus  rewrite $\delta J$ in terms of $\delta_m$ in the perturbed matter action. Finally, we can use the equation of motion for $\delta j$ 
\begin{equation}
\delta j = -\mathcal{N}_0 (\psi+v_m)\,
\end{equation}
in order to eliminate it in favour of $v_m$ and $\psi$.

Combining the above results and after some integrations by parts, we obtain the action for the scalar perturbations up to second order:
\begin{eqnarray}\label{MattAction}
  S_m^{(2)}&=&\int dt d^3x a^3
\Biggl[-{\frac {n\rho_{,{n}}(\partial v)^2}{2{a}^{2}}}+ \left( {\frac {3H \left( n{\rho_{,{n}}}^{2}-n\bar{\rho}\,\rho_{,{{nn}}}
-\bar{\rho}\,\rho_{,{n}} \right) \delta_{{m}}}{\rho_{,{n}}}}
+{\frac {n\rho_{,{n}} \partial^2\psi }{{a}^{2}}}-3\,n\rho_{,{n}}\dot{\zeta} -\bar{\rho}\,\dot{\delta}_m  \right) v_m \nonumber\\
&&{}-{\frac {\rho_{,{{nn}}}{\bar{\rho}}^{2}{\delta_{{m}}^2}}{2{\rho_{,{n}}^2}}}-\bar{\rho}\,\delta N \delta_{{m}}
\Biggr].
\end{eqnarray}
Notice that the velocity $v_m$ can always be integrated out, as $n\rho_{,n}=\bar{\rho}+p\neq0$.

\section{Study of Stability conditions}\label{Sec:stability}

In this section we present the main bulk of our work, i.e.\ the study of the general conditions that a  gravity theory has to satisfy in order to be free from instabilities when additional  matter fields are considered. 
These set of  requirements include:  no-ghost conditions,  positive  speeds of propagation (squared) and no-tachyonic instabilities~(see review \cite{Sbisa:2014pzo}).   
Recently, it has been shown that physical stability plays an important role when testing specific gravity models with cosmological data~\cite{Raveri:2014cka,Salvatelli:2016mgy}. In particular, the EFTCAMB patch~\cite{Hu:2013twa,Raveri:2014cka} includes a  specific module with the task to identify the viable parameter space of a selected theory. The results of the present work can be used to improve such modules and improve on the efficiency of the selection process.

To achieve this goal we consider the general EFT parametrization presented in section~\ref{Sec:EFT} in the presence of two different matter fluids, described by the action~(\ref{MattAction}),  for which we made the following, realistic, choices: a pressure-less fluid, i.e.\ cold dark matter/dust (d) and radiation (r).  A treatment which includes two general fluids complicates the process substantially and we do not expect to learn much more in such a case. So the relevant action required in  order to  proceed is of the following form:
\ba\label{FinalEFT}
S^{(2)}&=&\f{1}{(2\pi)^3}\int{}dtd^3ka^3 \left\{\bar{\rho}_d \left(-\frac{k^2 \psi}{a^2} -3   \dot{\zeta}-\dot{\delta}_d\right)v_d 
+\bar{\rho}_r\left(-\f{4}{3}\frac{k^2 \psi}{a^2}-4  \dot{\zeta}- \dot{\delta}_r\right)v_r 
-\bar{\rho}_d\frac{\left(k v_d\right)^2 }{2 a^2}-\f{2}{3}\bar{\rho}_r\frac{\left(k v_r\right)^2 }{ a^2}\r.\nn\\
&&\l.+\left(\frac{2 k^2 \zeta  \left(2 \hat{M}^2+m_0^2 (\Omega +1)\right)}{a^2}+3 F_2 \dot{\zeta}\right)\delta N + \left(\delta  N F_2-F_1 \dot{\zeta}\right)\frac{k^2 \psi }{a^2}-\frac{ F_4}{2 a^4}\left(k^2 \psi \right)^2+\frac{4m_2^2 (k\delta N )^2 }{a^2} \r.\nn\\
&&\l.+\frac{m_0^2 (\Omega +1) (k \zeta )^2}{a^2}-\f{4}{3}\bar{\rho}_r\frac{\left(k v_r\right)^2 }{2 a^2}- \bar{\rho}_d\delta N \delta_d+\frac{1}{2}  F_3\delta N^2-\frac{3}{2} F_1 \dot{\zeta}^2-\f{\bar{\rho}_r}{8}\delta _r^2 - \bar{\rho} _r\delta N \delta _r\right\},
\ea
where we have Fourier transformed the spatial coordinates and we have considered the following relations for the  number densities:
\be
n_d=\bar{\rho}_d\,, \qquad n_r=(\bar{\rho}_r)^{\f{3}{4}},
\ee
being $\bar{\rho}_d, \bar{\rho}_r$ respectively the density of dust and radiation at background. 

An action constructed in such a way admits only three DoFs described by $\{\zeta, \delta_d, \delta_r\}$. Therefore in the above action we  notice the presence of four Lagrange multipliers ${\delta N}$, $\psi$, $v_d$ and $v_r$.  Consequently, we proceed with the removal of the latter by using the constraint equations obtained after the variations of the action with respect  to the Lagrange multipliers. The resulting set of constraint equations is: 
\ba \label{constequations}
&&\bar{\rho}_r\left(-\f{4}{3}\frac{k^2 \psi}{a^2}-4  \dot{\zeta}- \dot{\delta}_r\right)-\f{4}{3}\bar{\rho}_r\frac{k^2 v_r }{ a^2}=0\,,\nn\\
&&\bar{\rho}_d \left(-\frac{k^2 \psi}{a^2} -3   \dot{\zeta}-\dot{\delta}_d\right)-\bar{\rho}_d\frac{k^2 v_d }{ a^2}=0\,,\nn \\
&&\frac{2 k^2 \zeta  \left(2 \hat{M}^2+m_0^2 (\Omega +1)\right)}{a^2}+3 F_2 \dot{\zeta}+\frac{8m_2^2 k^2\delta N  }{a^2}+  F_2\frac{k^2 \psi }{a^2}- \bar{\rho}_d \delta_d+ F_3\delta N- \bar{\rho} _r\delta _r=0\,,\nn\\
&&-\bar{\rho}_d v_d-\f{4}{3}\bar{\rho}_rv_r+\delta  N F_2-F_1 \dot{\zeta}-\frac{ F_4}{ a^2}k^2 \psi =0 \,.
\ea
After solving for the auxiliary fields and substituting the results back into action~(\ref{FinalEFT}), we get an action containing only the three dynamical DoFs $\{\zeta,\delta_d,\delta_r\}$:
\begin{equation}\label{actionshort}
S^{(2)}=\f{1}{(2\pi)^3}\int{} d^3kdta^3\l(\dot{\vec{\chi}}^t\textbf{A}
\dot{\vec{\chi}}-k^2\vec{\chi}^t\textbf{G}\vec{\chi}-
\dot{\vec{\chi}}^t\textbf{B}\vec{\chi}-\vec{\chi}^t\textbf{M}\vec{\chi}\r)\,,
\end{equation}
where we have defined the dimensionless vector:
\be
\vec{\chi}^t=(\zeta,\delta_d,\delta_r),
\ee
and the matrix components are listed in Appendix~\ref{APP:Coefficients}. In the next sections we will derive the stability conditions one needs to impose on the above action in order to guarantee the viability of the underlying theory of gravity.

Before proceeding with this in-depth analysis of the final action we present the background equations corresponding to our set-up: 
\begin{eqnarray} \label{backgroundeqs}
E_{1} & \equiv & 3m_0^2\left[1+\Omega+a\f{d\Omega}{da}\right]H^2+\Lambda-2\,c-\sum\bar{\rho}_i=0\,,\nn\\
E_{2} & \equiv & m_{0}^{2}(1+\Omega)(3H^{2}+2\dot{H})+2m_{0}^{2}\,H\dot{\Omega}+m_{0}^{2}\ddot{\Omega}+\sum_{i}p_{i}+\Lambda=0\,,\nn\\
E_{i} & \equiv & \dot{\bar{\rho}}_{i}+3H(\bar{\rho}_{i}+p_{i})=0\,.
\end{eqnarray}
where the Friedmann equations have been supplemented by the continuity equations for the fluids. 
 Finally, in order to close the system of equations, one needs to use the well-known equations of state for dust and radiation.  
As a side comment, from the background equations it is not possible to define in general a modified gravitational constant because $c$ and $\Lambda$ can be functions of $H^2$. The latter statement is clear when looking at the mapping of specific theories in the EFT language~\cite{Frusciante:2016xoj}.

\subsection{The presence of ghosts}\label{Sec:ghost}

A negative kinetic term of a field is usually considered as a pathology of the theory, since the high energy vacuum is unstable to the spontaneous production of particles~\cite{Carroll:2003st}. Such a pathology  must be constrained demanding for a positive kinetic term. 

Recently in ref.~\cite{Gumrukcuoglu:2016jbh}, it has been shown that such a constraint has to be imposed only in the high energy regime, in other words, an infrared ghost does not lead to a catastrophic vacuum collapse. On the contrary it was shown that it corresponds to a well known physical phenomenon, the Jeans instability.

In fact, expanding the ghost conditions in high-$k$ one can show that, when using appropriate field re-definitions, the sub-leading terms can be recast into the form of a Jeans mass instability, and viceversa. For example, the Hamiltonian $\mathcal{H}=-P^2+Q^2$ (where a ghost is present), can be recast into $\mathcal{H}=p^2-q^2$ (with negative squared speed of propagation and/or tachyonic mass), upon using the trivial canonical transformation $Q=p$, $P=-q$. Therefore, we will consider only the constraints coming from the high-$k$ behaviour  for the ghost  conditions as only in this regime they correspond to a true theoretical instability and not to a hidden physical phenomenon.  As for the tachyonic  squared mass (i.e.\ negative mass), it is problematic only when the time of evolution of the instability is much larger than $H^2$. We will elaborate on the latter in section~\ref{Sec:Mass}. 

Although the EFT approach has been discussed in the context of energies smaller than the cut-off of the theory, $\Lambda_{\rm cut-off}$,  here and in the following we will assume that we can still perform a high-$k$ expansion, namely we assume that in this regime we have $H\ll k/a\ll \Lambda_{\rm cut-off}$. This assumption is assumed to be valid at least for medium-low redshifts, those for which we can apply all the known cosmological-data constraints, namely BBN, CMB, BAO, etc.

In   action~(\ref{actionshort}) we have  a non-diagonal kinetic matrix for the three fields, i.e.\ $\mathcal{L}\ni A_{ij}\dot{\chi}_{i}\dot{\chi}_{j}$. As previously mentioned, in order to guarantee the absence of ghosts, one needs to demand  the high-$k$ limit of the kinetic matrix to be positive definite.
It is clear that one case encompassing all viable theories does not exist as a result of the wide range of operators which depend differently on the momentum. In particular, one has to pay attention to the operators accompanying $\bar{M}_2^3,\bar{M}^2_2$ and $m^2_2$, which exhibit a higher order dependence on $k$. Therefore, we will present a number of clear sub-cases which we consider relevant

We can identify a few cases:
\begin{enumerate}
\item  In this case all the functions in the Lagrangian are present, in particular $m_2^2\neq 0$ and $F_4\neq 0$. As a reference we note that the low-energy Ho\v rava gravity belongs to this general case. Expanding at high-$k$,  we find
  \begin{eqnarray}
   \mathcal{G}_1&=& {\frac { \left( F_{{1}}-3\,F_{{4}} \right) {a}^{3}F_{{1}}}{2F_4}}>0\,,\\
    \mathcal{G}_l&=& \frac{a^5\bar{\rho}_l^2}{2k^2 (\bar{\rho}_l+p_l)}>0\,,
  \end{eqnarray}
  where the index $l$ indicates the matter components, i.e.\ dust and radiation, $\mathcal{G}_1\equiv \mbox{Det}(\textbf{A})/(A_{22}A_{33}-A_{23}^2)$, $\mathcal{G}_r = A_{33}$ and $\mathcal{G}_d = A_{22}-A_{23}^2/A_{33}$. The $\mathcal{G}_l$ conditions represent the standard matter no-ghost conditions, which are trivially satisfied.
  \item  $F_4=0=m_2^2$. This case corresponds to the well known class of beyond Horndeski theories. We find:
  \begin{eqnarray}
    \mathcal{G}_1&=&{\frac { \left( F_{{1}}F_{{3}}+3\,{F_{{2}}}^{2} \right) F_{{1}}{a}^{3}}{2{F_{{2}}}^{2}}}>0\,,\\
    \mathcal{G}_l&=& \frac{a^5\bar{\rho}_l^2}{2k^2 (\bar{\rho}_l+p_l)}>0\,.
  \end{eqnarray}
\item  $F_4= 0$ and $m_2^2\neq 0$. The ghost conditions change into
  \begin{eqnarray}
    \mathcal{G}_1&=&{\frac {4F_{{1}}^{2}m_{{2}}^{2}{k}^{2}a}{F_{{2}}^{2}}}>0\,,\\
    \mathcal{G}_d&=& {\frac {F_{{2}}^{2}{a}^{5}{\bar\rho}_d}{2{k}^{2}
        (F_{{2}}^{2} -8\,m_{{2}}^{2}{\bar\rho}_d) }}>0\,,\\
    \mathcal{G}_r&=&
    {\frac {9\,{a}^{5} \left( F_{{2}}^{2}-8\,m_{{2}}^{2}{\bar\rho}_{{d}}
 \right) {\bar\rho}_{{r}}}{8\,{k}^{2} [ 3\,F_{{2}}^{2} -8\,m_{{2}}^{2}(3{\bar\rho}_{{d}}+4{\bar\rho}_{{r}}) ] }}>0\,,
  \end{eqnarray}
  and in this case the matter no-ghost conditions get non-trivially modified. In particular, we find $0<m_2^2<F_2^2/[8({\bar\rho}_d+4{\bar\rho}_r/3)]$. 
	Such condition prevents  $m_2^2$ to be arbitrarily large ensuring the stability of the theory. One might wondering about  the role of spatial gradients of the lapse in the stability of matter, since in the action (\ref{FinalEFT}) there is no direct coupling between matter and gravity.  However, the spatial gradient of the lapse turns out to be proportional to $\dot{\delta}_d^2$, $\dot{\delta}_r^2$ and $\dot{\zeta}^2$ through eqs. (\ref{constequations}), then it is directly involved in the above ghost conditions.  In this sense there is a "coupling" between gravity and matter fields.	
\item  $m_2^2=0$ and $F_4\neq 0$. In this case we have
  \begin{eqnarray}
    \mathcal{G}_1&=&{\frac {{a}^{3} \left( F_{{1}} - 3\,F_4\right)  \left( F_{{1}
}F_{{3}}+3\,F_{{2}}^{2} \right) }{2({F_{{2}}}^{2}+F_4F_{{3}})}}>0\,,\\
    \mathcal{G}_l&=& \frac{a^5\bar{\rho}_l^2}{2k^2 (\bar{\rho}_l+p_l)}>0\,.
  \end{eqnarray}
\item  $F_1=0$. In this case the no-ghost conditions can be written as:
  \begin{eqnarray}
    \mathcal{G}_1&=&-{\frac {9\,F_{{2}}^{2}{a}^{5}}{16\,m_{{2}}^{2}{k}^{2}}}>0\,,\\
    \mathcal{G}_l&=& \frac{a^5\bar{\rho}_l^2}{2k^2 (\bar{\rho}_l+p_l)}>0\,,
  \end{eqnarray}
  so that $m_2^2<0$.
\item Cases: $F_1=3F_4$, or $F_1=0=F_2$, or $m_2^2=0=F_1F_3+3F_2^2$. In this cases the determinant of the kinetic matrix identically vanishes. This behaviour, in general, leads to strong coupling, so that this class of theories cannot be considered as valid EFT.
\end{enumerate}

A final remark on the first two cases, which are the most noticeable since they are strictly related to well known models: the presence of matter fluids does not affect the form of the ghost conditions, indeed, we recover the same results as in ref.~\cite{Frusciante:2016xoj} where no matter fluids were included,  once the high-$k$ limit has been taken. 
However, let us note that the parameters space identified by these conditions can change because of the evolution of the scale factor, which in turns is the solution of different Friedmann equations.  Moreover, no-ghost conditions have been previously obtained in presence of matter fields described by a $P(\mathcal{X})$ action as in refs.~\cite{Kase:2014cwa,DeFelice:2011bh,Gleyzes:2015pma,D'Amico:2016ltd} (and references therein). Such results are obtained for the variable $v_m$ and they can be safely applied for all matter fluids but not for dust. Indeed, in the specific case of pressureless fluids ($w\rightarrow 0$) the ghost condition turns out to be ill defined. This can be explained by the fact that the no-ghost conditions need to be derived  at the level of the action,  which diverges in this limit. From a physical point of view this is related to a ''bad'' choice of  physical variable which has to describe the matter DoFs as we discussed in section~\ref{Sec:matter}.  However, in some cases they can be extended to non relativistic matter species as for eg. in ref.~\cite{DeFelice:2011bh},  where the authors use for the barotropic coefficient of these species the case $w = 0^+$ which implies a small yet non-negligible pressure and  speed of propagation. In conclusion, by using appropriate precautions in some cases present in literature one can find some of the above results, mostly related to case 2. In this sense our results are more general and robust.

\subsection{The speeds of propagation}\label{Sec:speed}

We will now proceed with the study of the speeds of propagation associated to the scalar DoFs in action~(\ref{actionshort}). As usual, their positivity guarantees the avoidance of any potential gradient instabilities at high-$k$. Hereafter, we will consider the action purely in the high-$k$ limit. This is a necessary step in order to obtain the physical speeds of propagation. Indeed,  if one does not assume the high-$k$ limit the resulting "speeds of propagation" would be  complicated and non-local expressions due to the complex dependence on the momentum of the action~(\ref{actionshort}) and the interaction between the three fields. Of course, in order to  study  the gradient instability in full generality one needs to work out such expressions. However, let us say that in such case the fields do not decouple from each other and it turns out to be very difficult  to obtain analytical expressions for the speeds of propagation.  Moreover, the regime in which the gradient instability manifests itself faster and thus becomes  potentially dangerous within the lifetime of the universe is in the high-$k$ limit, thus justifying our restriction to such a regime. 

In order to achieve this,  it is necessary to diagonalise the kinetic matrix, therefore we will proceed with the following field redefinition:
\begin{eqnarray}\label{diagonal}
\zeta & = & \Psi_{1},\qquad
\delta_{d} = \Psi_{2}k-\frac{A_{12}A_{33}-A_{13}A_{23}}{A_{22}A_{33}-A_{23}^{2}}\,\Psi_{1},\qquad
\delta_{r} = k\Psi_{3}+\frac{A_{12}A_{23}-A_{13}A_{22}}{A_{22}A_{33}-A_{23}^{2}}\,\Psi_{1}-\frac{A_{23}}{A_{33}}\,\Psi_{2}k\,.
\end{eqnarray}
The $k$ dependence of the transformation is a convenient choice in order to obtain, in the high-$k$ limit, a scale invariant kinetic matrix and the new kinetic matrix, $\mathcal{L}\ni a^3 K_{ij}{\dot\Psi}_i{\dot\Psi}_j$,  is now diagonal without approximations. Finally, we get a Lagrangian of the form:
\begin{equation}
\mathcal{L}^{(2)}=K_{11}\dot{\Psi}_{1}^{2}+K_{22}\dot{\Psi}_{2}^{2}+K_{33}\dot{\Psi}_{3}^{2}+Q_{12}(\dot{\Psi}_{1}\Psi_{2}-\dot{\Psi}_{2}\Psi_{1})+Q_{13}(\dot{\Psi}_{1}\Psi_{3}-\dot{\Psi}_{3}\Psi_{1})+Q_{23}(\dot{\Psi}_{2}\Psi_{3}-\dot{\Psi}_{3}\Psi_{2})-\mathcal{M}_{ij}\Psi_{i}\Psi_{j}\,,
\end{equation}
where the kinetic matrix coefficients are:
\ba
&&K_{11}= \frac{A_{33} A_{12}{}^2-2 A_{13} A_{23} A_{12}+A_{13}^2 A_{22}+A_{11} \left(A_{23}{}^2-A_{22} A_{33}\right)}{A_{23}{}^2-A_{22} A_{33}}\,,\nn\\
&&K_{22}= k^2\l(A_{22}-\frac{A_{23}{}^2}{A_{33}}\r)\,,\nn\\
&&K_{33}=k^2A_{33} \qquad K_{ij}=0\,\,\ \mbox{with}\,\,\, i\neq j\,,
\ea
and  the $Q_{ij}$ and $\mathcal{M}_{ij}$ matrix coefficients will be specified in the following case by case.

Due to the different scaling with k of the operators in action~(\ref{actionshort}), it is necessary to analyse the sub-cases identified before separately. As it will become clear every sub-case exhibits a different behaviour, as expected. 
\begin{enumerate}
\item  General case ($m_2^2\neq 0$ and $F_4\neq 0$). The kinetic matrix elements at high-$k$ read
\begin{equation}
K_{11}=\frac{F_1 \left(F_1-3 F_4\right)}{2 F_4}+ \mathcal{O}(k^{-2})\,,\qquad K_{22}=\frac{1}{2} a^2 \bar{\rho}_d+ \mathcal{O}(k^{-2})\,,\qquad K_{33}=\frac{3}{8} a^2 \bar{\rho}_r+ \mathcal{O}(k^{-2})\,,
\end{equation}
which are scale invariant. In its full generality, the action reduces in such a limit to a  system of three decoupled fields:
\ba
S^{(2)}&=&\f{1}{(2\pi)^3}\int{}dk^3dt\f{a^3}{8}\l\{4 a^2 \bar{\rho}_d \dot{\Psi}_2^2+3 a^2 \bar{\rho} _r \dot{\Psi}_3^2+\frac{4 F_1 \left(F_1-3 F_4\right) }{F_4}\dot{\Psi}_1^2\r.\nn\\
&&\l.-\f{k^2}{a^2}\l[ \frac{ 2\left(-4 m_0^2 (\Omega +1) \left(m_2^2-\hat{M}^2\right)+4 \hat{M}^4+m_0^4 (\Omega +1)^2\right)}{  m_2^2}\Psi _1^2+a^2\bar{\rho} _r\Psi_3^2\r]+ \mathcal{O}(k^{-1})\r\},
\ea
from which it is easy to read off the $Q_{ij}$ and $\mathcal{M}_{ij}$ coefficients. Then, for high-$k$, the elements $Q_{ij}$ are corrections and the
matrix $\mathcal{M}_{ij}$ becomes diagonal. This decoupling is very helpful when obtaining the speeds of propagation from the Euler-Lagrange equations:
\ba
&& \frac{F_1 \left(F_1-3 F_4\right)}{F_4}\ddot{\Psi}_1+ \f{k^2}{a^2}\l(\frac{-4 m_0^2 (\Omega +1) \left(m_2^2-\hat{M}^2\right)+4 \hat{M}^4+m_0^4 (\Omega +1)^2}{2  m_2^2}\r)\Psi_1\nn\\
&&\,\,\,\, +\l(\frac{F_1^2 \left(3 F_4 H- \dot{F}_4\right)+F_1 F_4 \left(2  \dot{F}_1-9 F_4 H\right)}{F_4^2}-3 \dot{F}_1\r)\dot{\Psi}_1 \approx 0 \,,\nn\\
&&\ddot{\Psi}_2+2 H \dot{\Psi}_2 \approx 0\,, \nn\\
&&3\ddot{\Psi}_3+3 H \dot{\Psi}_3+\f{k^2}{a^2} \Psi _3 \approx 0.
\ea
It is now straightforward to isolate the three speeds of propagation and look at their functional dependence:
\ba\label{speeds}
c^2_{s,g}= \frac{F_4 \left(-4 m_0^2 (\Omega +1) \left(m_2^2-\hat{M}^2\right)+4 \hat{M}^4+m_0^4 (\Omega +1)^2\right)}{2 F_1 m_2^2 \left(F_1-3 F_4\right)}\,,\qquad c^2_{s,d}=0 \,,\qquad c^2_{s,r}=\f{1}{3},
\ea
where we have used the suffix $''g''$ to indicate the speed of propagation  associated to the DoF of the gravity sector.
It is clear that when we consider all the operators active, including the higher order in spatial derivative operators, one gets a completely decoupled system where the fields do not influence each other and evolve separately. 
\item Case $F_4=0=m^2_2$. After applying the fields re-definitions~(\ref{diagonal}), we get in the large k-limit the following action:
\ba
S^{(2)}&=&\f{1}{(2\pi)^3}\int{}dk^3dt a^3\l\{\frac{1}{2} F_1 \left(\frac{F_1 F_3}{F_2^2}+3\right) \dot{\Psi}_1^2+ \frac{1}{2} a^2 \bar{\rho}_d \dot{\Psi}_2^2+ \frac{3}{8 }a^2 \bar{\rho}_r\dot{\Psi}_3^2-k^2\frac{ \bar{\rho}_r}{8 }\Psi_3^2 \r.\nn\\
&&\l.+ k\frac{1}{2  F_2} \left(4 \hat{M}^2+2 m_0^2 (\Omega+1)-F_1\right)\l[\bar{\rho}_d\l(\Psi _2 \dot{\Psi}_1-\Psi_1 \dot{\Psi}_2\r)+\bar{\rho}_r\l(\Psi _3 \dot{\Psi}_1-\Psi_1 \dot{\Psi}_3\r)\r]\r.\nn\\
&&\l.+ \frac{k^2}{3 a^2 F_2^2}\left[-3  F_1 F_2 H \left(2 \hat{M}^2+m_0^2 (\Omega +1)\right)+ \l(6  \bar{\rho}_d+8 \bar{\rho}_r\r) \left(2 \hat{M}^2+m_0^2 (\Omega +1)\right)^2+3 m_0^2 \left[F_1 (\Omega +1) \dot{F}_2 \r.\r.\r. \nn\\
&&\l.\l.\l.-F_2 \left(F_1 \dot{\Omega}+(\Omega +1) \dot{F}_1\right)+F_2^2 (\Omega +1)\right]-6 \left(F_2 \left(\dot{F}_1 \hat{M}^2+2F_1 \hat{M}\dot{\hat{M}}\right)-F_1 \dot{F}_2 \hat{M}^2\right)\right]\Psi_1^2\r\}+\mathcal{O}(k^{-2}).\nn\\
&&
\ea 

As it is clear, the resulting action in the high-$k$ limit exhibits some substantial deviations from the previous case. 
The complication arises due to the fact that now the fields are coupled in antisymmetric configurations. This will force us to change approach when obtaining the speeds of propagation. Namely, we will choose  firstly to Fourier transform the time component in the Lagrangian by using ($\partial_t \rightarrow -i\omega$) and then proceed to obtain the dispersion relations. This will yield the following:
\be
\mathcal{L}^{(2)}\sim \l(\Psi_1, \Psi_2, \Psi_3\r)\left(
\begin{array}{ccc}
\frac{1}{2} F_1 \left(\frac{F_3 F_1}{F_2^2}+3\right)\omega^2-\f{k^2}{ a^2}\mathcal{G}_{11} & - i \omega k \mathcal{B}_{12} &  - i \omega k \mathcal{B}_{13} \\
 i \omega k \mathcal{B}_{12} & \frac{1}{2} a^2 \bar{\rho}_d \omega^2 & 0\\
i \omega k \mathcal{B}_{13} & 0 & \frac{3}{8 }a^2 \bar{\rho}_r\omega^2-k^2\frac{ \bar{\rho}_r}{8 }
\end{array}
\right)\left( \begin{array}{cc} &\Psi_1 \\ &\Psi_2 \\& \Psi_3
\end{array}
\right)\,,
\ee  
where $\mathcal{G}_{11}$ and $\mathcal{B}_{ij} $  can be read off from the action. Now, setting the determinant of the above matrix to zero and  considering that $\omega^2=\f{k^2}{a^2}c_s^2$ in the high-$k$ limit,  we obtain the following results: 
\ba\label{beyondspeed}
&&c_{s,d}^2=0\,,\nn\\
&& (3 c^2_s-1) \bar{\rho} _r \left[\bar{\rho} _d \left(c^2_s (F_3 F_1^2+3 F_2^2 F_1)-2 a^2 F_2^2 \mathcal{G}_{11}\right)-4 \mathcal{B}_{12}^2 F_2{}^2\right]-16 c^2_s \mathcal{B}_{13}^2 F_2^2 \bar{\rho} _d=0
\ea
with $F_2\neq 0$ and where $c_s^2$ is the double solution of the dispersion relation obtained after observing that the dust speed of propagation, $c_{s,d}^2$ is zero. It is clear that, while the speed of propagation of the dust component remains unaffected by the presence of radiation and gravity, the last dispersion relation manifests the clear interaction between radiation and gravity, which  modifies their speeds of propagation. Hence, this result shows us clearly that the interaction with matter can affect the gravity sector in a very deep way. 

The only case in which the gravity sector and the radiation one completely decouple is when the following condition applies: 
\be \label{condition}
4 \hat{M}^2+2 m_0^2 (\Omega+1)-F_1=0.
\ee
In this case from~(\ref{beyondspeed}) the standard speed of propagation for the radiation is recovered and the speed of gravity is
\be \label{speedgravity}
c^2_{s,g} = \f{ 2 F_2^2 \mathcal{G}_{11}}{F_3 F_1^2+3 F_2^2 F_1}\,.
\ee
Let us notice that the condition~(\ref{condition}) is trivially satisfied for the Horndeski class of models. In Eq.~(\ref{speedgravity}), $\mathcal{G}_{11}$ depends on the background densities of dust and radiation, then one can use the background equations~(\ref{backgroundeqs}) to eliminate the dependence from the densities of the matter fluids, thus obtaining 
\be\label{speedfinal}
c^2_{s,g}=\frac{2 \left(2 c F_1^2+2 m_0^2 F_1^2  \dot{H}(\Omega+1)+F_1^2 H \left(F_2-m_0^2 \dot{\Omega} \right)+m_0^2 F_1^2 \ddot{\Omega}-2 m_0^2 F_2^2 (\Omega +1)-F_1^2 \dot{F}_2+2 F_2 F_1 \dot{F}_1\right)}{F_1 \left(3 F_2^2+F_1 F_3\right)}.
\ee
Even though the radiation and the dust sector appear unaltered there is some interplay between gravity and the matter sector.  Although the above expression for the speed of propagation of the gravity mode holds both in the vacuum and matter case, the parameters space defined through Eq.~(\ref{speedfinal}) changes drastically in the two cases. Indeed, firstly one has to consider a different evolution for the scale factor, $a(t)$ accordingly to the corresponding  Friedmann equations, secondly in the vacuum case Eq.~(\ref{speedfinal}) simplifies because a combination of terms turns to be zero due to the Friedmann equations. Instead, such combination of terms when matter is included  gives a non zero contribution.

The same result for this sub-case has been obtained in ref.~\cite{Kase:2014cwa,D'Amico:2016ltd}, starting from a $P(\mathcal{X})$ action for the matter sector and the $v_m$ variable. It is important to note that, in contrast to the no-ghost conditions, the results also agree for the case of dust. This can be explained by the fact that the speed of propagation can be obtained at the level of the equations of motion, hence avoiding the issues plaguing the action, described in the previous sections.

\item Case  $F_4 = 0$ and $m_2^2\neq 0$. The action at high-$k$ reads
\ba
S^{(2)}&=& \f{1}{(2\pi)^3}\int{}dk^3dta^3\l\{ \frac{4 k^2 F_1^2 m_2^2}{a^2F_2^2}\dot{\Psi}_1^2+\frac{a^2 \bar{\rho}_{d} F_2^2}{2 F_2^2-16 \bar{\rho}_{d} m_2^2} \dot{\Psi}_2^2+\frac{9 a^2 \bar{\rho} _{r} \left( F_2^2-8 \bar{\rho}_{d} m_2^2\right)}{8 \left(-24 \bar{\rho}_{d} m_2^2+3  F_2^2-32 m_2^2 \bar{\rho}_{r}\right)}\dot{\Psi}_3^2-\frac{k^2 \bar{\rho}_r}{8}\Psi_3^2 \nn\r.\\
&&\l. -\frac{128 k^2 \bar{\rho} _{d}^2 m_2^4 \bar{\rho} _r}{9  \left( F_2^2-8 \bar{\rho} _{d} m_2^2\right)^2}\Psi_2^2-\frac{128 k^4 F_1^2 m_2^4 \bar{\rho} _{r}}{9 a^4 F_2^4}\Psi_1^2 +\frac{256 k^3 a\bar{\rho} _{d} F_1 m_2^4 \bar{\rho}_{r}}{9 a^3 F_2^4-72  \bar{\rho} _{d} F_2^2 m_2^2}\Psi_1\Psi_2-\frac{8 k^3 F_1 m_2^2 \bar{\rho}_r}{3 a^2 F_2^2}\Psi_1\Psi_3\r.\nn\\
&&\l. +k \l(16 F_1 F_2 m_2^2 H+\left(4 F_2 \left(F_2 \hat{M}^2-2 m_2^2 \dot{F}_1\right)+2 m_0^2 F_2^2 (\Omega+1)-F_1 \left(16 F_2 m_2\dot{m}_2-16 m_2^2 \dot{F}_2+F_2^2\right)\right)\r)\times\r.\nn\\
&&\l.\times \l[\f{\bar{\rho}_d}{2 \left( F_2^3-8 \bar{\rho} _{d} F_2 m_2^2\right)}\l(\Psi _2 \dot{\Psi}_1-\Psi _1 \dot{\Psi}_2\r)+\f{3\bar{\rho}_r}{2 F_2 \left(-24  \bar{\rho} _{d} m_2^2+3  F_2^2-32 m_2^2 \bar{\rho} _{r}\right)}\l(\Psi _3\dot{\Psi} _1-\Psi _1\dot{\Psi} _3\r)\r] \r.\nn\\
&&\l.+\frac{8 k^2 \bar{\rho} _{d} m_2^2 \bar{\rho}_r}{3  F_2^2-24 \bar{\rho}_{d} m_2^2}\Psi_2\Psi_3   
\r\}+O(k^{-2})\,.
\ea
We find that the solutions of the discriminant equation
\be
\det\!\left(\frac{c_{2}^{2}k^{2}}{a^{2}}\,K_{ij}-\mathcal{M}_{ij}\right)=
{\frac {{a}^{5}{c_{{s}}}^{4}\bar{\rho}_{{m}} \left( {c_{{s}}}^{2}\rho_{{r,n}
}-n_{{r}}\rho_{{r,{\it nn}}} \right) {\bar{\rho}_{{r}}}^{2}{m_{{2}}}^{2}{F_{
{1}}}^{2}{k}^{6}}{ \left( -8\,{m_{{2}}}^{2}n_{{r}}\rho_{{r,n}}-8\,{m_{
{2}}}^{2}\bar{\rho}_{{m}}+{F_{{2}}}^{2} \right) n_{{r}}{\rho_{{r,n}}}^{2}}}\,,
\ee
 reduce to
\begin{equation}
c_{s,g}^{2}=0\,,\qquad 
c_{s,d}^{2}  =0\,,\qquad
c_{s,r}^{2}   =\frac13\,.
\end{equation}
The results for this case can be found in the limit $F_{4}\to0$ for the general case discussed above.

\item Case $F_4\neq 0$ and $m_2^2=0$. The action for this sub-case at high-$k$ reads
\ba
S^{(2)}&=& \f{1}{(2\pi)^3}\int{}dk^3dta^3\l\{\frac{ \left(3 F_2^2+F_1 F_3\right) \left(F_1-3 F_4\right)}{2 \left(F_2^2+F_3F_4\right)}\dot{\Psi}_1^2+\frac{1}{2} a^2 \bar{\rho}_{d}\dot{\Psi}_2^2+\frac{3}{8} a^2 \bar{\rho} _{r}\dot{\Psi}_3^2-k^2\frac{\bar{\rho}_r \left(F_2^2+F_3 F_4+4 F_4 \bar{\rho}_{r}\right)}{8 \left(F_2^2+F_3 F_4\right)}\Psi_3^2\r. \nn\\
&&\l.-\frac{k^2\bar{\rho}_{d}^2 F_4}{2 \left(F_2^2+F_3 F_4\right)} \Psi_2^2-k^4\frac{2  F_4 \left(2 \hat{M}^2+m_0^2 (\Omega+1)\right)^2}{a^4 \left(F_2^2+F_3 F_4\right)}\Psi_1^2 +k^3\frac{2 F_4  \left(2 \hat{M}^2+m_0^2 (\Omega+1)\right)}{a^2 \left(F_2^2+F_3 F_4\right)}  \l(\bar{\rho}_r \Psi_3\Psi_1+\bar{\rho}_d\Psi_1\Psi_2\r)  \r.\nn\\
&&\l.+k\frac{ F_2 \left(-F_1+3 F_4+4 \hat{M}^2+2 m_0^2 (\Omega+1)\right)}{2 \left(F_2^2+F_3F_4\right)}\l[\bar{\rho}_{d}\l(\Psi_2 \dot{\Psi}_1-\Psi_1 \dot{\Psi}_2\r)+\bar{\rho}_r\l(\Psi_3 \dot{\Psi}_1-\Psi_1 \dot{\Psi}_3\r)\r]\r.\nn\\
&&\l.-\frac{k^2 \bar{\rho}_{d} F_4 \bar{\rho}_r}{\left(F_2^2+F_3 F_4\right)}\Psi_2\Psi_3\r\}+\mathcal{O}(k^{-2}) \,,
\ea
where the kinetic terms $K_{11},K_{22},K_{33}$ are of order $\mathcal{O}(k^0)$ for high values of $k$ and the elements $Q_{12}$
and $Q_{13}$ are of order $k$ and cannot be neglected. Furthermore, the leading component of $\mathcal{M}_{11}$ is of order $k^{4}$. 
Therefore now we need to consider the discriminant equation as
\begin{equation}
\mbox{Det}(\omega^{2}K_{ij}-i\,\omega\,Q_{ij}-\mathcal{M}_{ij})=0\,,
\end{equation}
this equation can be recast as
\be
\omega^{6}+\left(\mathcal{A}\,\frac{k^{4}}{a^{4}}+\mathcal{O}(k^{2})\right)\omega^{4}+\left(\mathcal{B}\,\frac{k^{6}}{a^{6}}+\mathcal{O}(k^{4})\right)\omega^{2}=0\,,\label{eq:discrim20}
\ee
with
\ba
\mathcal{A}=4\,\frac { \left( (\Omega+1)\,{m_{{0}}}^{2}+2\,\hat{M}^{2} \right)^{2}F_{4}}{\left( 3\,F_{{4}}-F_{{1}}
 \right)  \left( F_{{1}}F_{{3}}+3\,{F_{{2}}}^{2} \right) }\,, \qquad \mathcal{B}=-\frac{1}{3}\,\mathcal{A}\,.
\end{eqnarray}
For high-$k$, we find the following solutions:
\begin{itemize}
\item One solution can be found by assuming $\omega^{2}=W\,k^{4}/a^{4}$. In this case we find
\be
(W+\mathcal{A})\,W^{2}\,\frac{k^{12}}{a^{12}}+\mathcal{O}(k^{10})=0\,,
\ee
which is verified by $W=-\mathcal{A}$, so that
\begin{equation}
\omega^{2}=-\mathcal{A}\,\frac{k^{4}}{a^{4}}\,,\qquad c_{s,g}^2=-4\mathcal{A}\,\frac{k^{2}}{a^{2}}\,,
\end{equation}
or
\begin{equation}
c_{s,g}^{2}=\frac{16F_{{4}}\left((\Omega+1)\,m_{{0}}^{2}+2\,\hat{M}^{2}\right)^{2}}{\left(F_{{1}}F_{{3}}+3\,F_{{2}}^{2}\right)\left(F_{{1}}-3F_{4}\right)}\,\frac{k^{2}}{a^{2}}\,,
\end{equation}
which tends to large values.
\item The other two solutions of Eq. (\ref{eq:discrim20}) can be found by
assuming $\omega^{2}=Wk^{2}/a^{2}$, so that
\begin{equation}
\left(\mathcal{A}\,W^{2}+\mathcal{B}W\right)\,\frac{k^{8}}{a^{8}}+\mathcal{O}(k^{6})=0\,,
\end{equation}
which implies the following standard results
\begin{equation}
c_{s,d}^{2}  =  0\,,\qquad c_{s,r}^{2}  =  \frac{1}{3}\,.
\end{equation}
\end{itemize}

\item Case $F_1=0$. The action reads:
\ba
S^{(2)}&=&\f{1}{(2\pi)^3}\int{}dk^3dt a^3\l[-\frac{9 a^2 F_2^2}{16 m_2^2}\dot{\Psi}_1^2+\frac{\bar{\rho}_d}{2}a^2\dot{\Psi}_2^2+\f{3}{8}\bar{\rho}_ra^2\dot{\Psi}_3^2-\frac{k^2 \bar{\rho}_r}{8}\psi_3^2+k^2\frac{ \bar{\rho}_d\left(2 \hat{M}^2+m_0^2 (\Omega+1)\right)}{4  m_2^2}\psi_1\Psi_2\r.\nn\\
&&\l. +k^2\frac{ \bar{\rho}_r\left(4 \hat{M}^2+2 m_0^2 (\Omega+1)+8 m_2^2\right)}{8  m_2^2}\Psi_1\Psi_3-k^4\frac{4 \hat{M}^4-4 m_0^2 (\Omega+1) \left(m_2^2-\hat{M}^2\right)+m_0^4 (\Omega+1)^2}{4 a^2 m_2^2}\Psi_1^2 \r] \nn\\
&&+\mathcal{O}(k^{-2}).
\ea
In this case, the matrix $Q_{ij}$ can be neglected, but the $\mathcal{M}_{11}$ coefficient has a term in $k^{4}$, therefore we have the discriminant equation
\be
\mbox{Det}(\omega^{2}K_{ij}-\mathcal{M}_{ij})=0\,,
\ee
which leads to
\begin{equation}
\omega^{6}+\left(\mathcal{A}\,\frac{k^{4}}{a^{4}}+\mathcal{O}(k^{2})\right)\omega^{4}+\left(-\f{1}{3}\mathcal{A}\,\frac{k^{6}}{a^{6}}+\mathcal{O}(k^{4})\right)\omega^{2}+\mathcal{O}(k^{6})=0\,.\label{eq:discriF10}
\end{equation}
with
\be
\mathcal{A}= \frac{4 \left(-4 m_0^2 (\Omega+1) \left(m_2^2-\hat{M}^2\right)+4 \hat{M}^4+m_0^4 (\Omega+1)^2\right)}{9 F_2^2}\,.
\ee
Once more we have a solution
\begin{equation}
\omega^{2}=-\mathcal{A}\,\frac{k^{4}}{a^{4}}\,,
\end{equation}
which leads to
\begin{equation}
c_{s,g}^{2}=\frac{16}{9}\,\frac{4\,(m_{{2}}^{2}-\hat{M}^{2})\left(1+\Omega\right)m_{0}^{2}-\left(1+\Omega\right)^{2}m_{0}^{4}-4\,\hat{M}^{4}}{F_{{2}}^{2}}\,\frac{k^{2}}{a^{2}}\,,
\end{equation}
whereas the other two solutions are found to be 
\begin{equation}
c_{s,d}^{2} = 0\,,\qquad
c_{s,r}^{2}  = \f{1}{3}\,.
\end{equation}
 \end{enumerate}

In summary, in this section we have derived the speeds of propagation for the three dynamical fields describing our system. In general for all the sub-cases analysed we have found that in the high-$k$ regime the three DoFs decouple and the resulting speeds of propagation are unaltered with respect to the vacuum case. This can be easily  verified by considering the high-$k$ limit of the results in ref.~\cite{Frusciante:2016xoj}. Only one case stands aside, the beyond Horndeski case. In this case the dust field completely decouples from the other fields, while the radiation and gravity fields are coupled and their speeds result to be modified. 
We also recall that even in the cases the expressions for the speeds of propagation do not differ from the respective  cases in vacuum, the parameters space may change accordingly to a different evolution of the scale factor, which in turns is the solution of different background equations. 
In conclusion, for all the cases analysed we demand a positive speed of propagation in order to guarantee the viability of the underlying  theory of gravity.

\subsection{Tachyonic  and Jeans instabilities}\label{Sec:Mass}

The final aspect of our work which tends to be the one least studied in the literature in the context of MG theories and especially in the EFT framework, is the study of the canonical mass of the fields and consequently the boundedness of the Hamiltonian at low momenta~\cite{Frusciante:2016xoj}. These results are related to the usual tachyonic and  Jeans instabilities, the latter being characteristic of the fluids components.

In this section we will restrict the analysis to the EFT action in the presence of only one matter fluid. We choose the dust over radiation because we know that the dust component clusters and hence for our purpose it might show an interesting behaviour related to the  Jeans instability.  Thus, the results presented here will be applicable during the dust-dominated era and onwards when the dynamics of the two  DoFs starts to play a role.  A second fluid can be straightforwardly added, but it makes the  procedure substantially more difficult.  

One can obtain the action for the EFT with a dust component by setting $\delta_r=0$ in the action~\eqref{FinalEFT} and $\bar{\rho}_r=0$ in the remaining functions. Now, let us  assume the no-ghost conditions hold, and  proceed to rewrite the action in its canonical form.
The  first step is to diagonalise the (2x2) kinetic matrix as in the previous section by making the following field redefinitions 
\ba
&&\zeta=\Psi_1\,,\nn\\
&&\delta_d=k \Psi _2-\frac{A_{12} \Psi _1}{A_{22}}\,,
\ea
with the following diagonal terms:
\be
\bar{K}_{11}= A_{11}-\frac{A_{12}^2}{A_{22}}, \qquad \bar{K}_{22}=k^2 A_{22}\,,
\ee
where $A_{ij}$ are the ones defined in Appendix~\ref{APP:Coefficients} after setting $\bar{\rho}_r=0$. Next, the canonical form is obtained by normalising the fields accordingly to:
\begin{eqnarray}
\Psi_1 &=& \frac{1}{\sqrt{2\bar{K}_{11}}}\,\bar\Psi_1\,, \nn\\
\Psi_2 &=& \frac{1}{\sqrt{2\bar{K}_{22}}}\,\bar\Psi_2\,.
\end{eqnarray}
After grouping the different terms and performing a number of integrations by parts, we obtain the Lagrangian as:
\begin{equation}\label{actionmass}
  \mathcal{L}^{(2)}=\frac{a^3}2\left[\dot{\bar\Psi}_1^2 +\dot{\bar\Psi}_2^2 + \bar B(t,k)\,(\dot{\bar\Psi}_1{\bar\Psi}_2-\dot{\bar\Psi}_2{\bar\Psi}_1)-\bar C_{ij}(t,k){\bar\Psi}_i{\bar\Psi}_j\right],
\end{equation}
where we refer the reader to the Appendix~\ref{APP:Coefficients} for the functional forms of the $\bar{B}, C_{ij}$ coefficients.\\
In order to obtain the mass eigenvalues we need to proceed with the diagonalization of the mass matrix $C_{ij}$ while keeping the canonical form of the action.
For this purpose we consider a field rotation via an orthogonal matrix,  in the following way:
\begin{eqnarray}
\bar\Psi_1&=&\cos\alpha\,\Phi_1+\sin\alpha\,\Phi_2\,, \nn\\
\bar\Psi_2&=&-\sin\alpha\,\Phi_1+\cos\alpha\,\Phi_2\,.
\end{eqnarray}
Now, it is possible to choose $\alpha$ in a very specific way in order to diagonalize the mass matrix. This leads to the following relation:
\begin{equation}
\tan(2\alpha)=\beta\equiv-\frac{2\bar C_{12}}{\bar C_{11}-\bar C_{22}}\,,
\end{equation}
accompanied by:
\begin{equation}
\begin{aligned}[c]
\frac{d[\tan(2\alpha)]}{dt}=2[1+\tan^2(2\alpha)]\dot\alpha
\end{aligned}
\qquad\Longleftrightarrow\qquad
\begin{aligned}[c]
\dot\alpha=\frac{\dot\beta}{2(1+\beta^2)}\,.
\end{aligned}
\end{equation}
Then, the Lagrangian becomes
\begin{equation}
\label{finalLagr}
\mathcal{L}^{(2)} = \frac{a^3}2\left[\dot{\Phi}_1^2 +\dot{\Phi}_2^2 + B(t,k)\,(\dot{\Phi}_1{\Phi}_2-\dot{\Phi}_2{\Phi}_1)-\mu_1(t,k)\Phi_1^2
-\mu_2(t,k)\Phi_2^2\right],
\end{equation}
with the following definitions:
\begin{eqnarray}
B&=&\bar B+2\dot\alpha\,, \nn\\
\mu_1&=&-\dot\alpha^2-\bar B\dot\alpha+\frac{(\bar C_{11}-\bar C_{22})^2+4C_{12}^2}{\bar C_{11}-\bar C_{22}}\,\cos^2\alpha
+\frac{\bar C_{11}\bar C_{22}-2\bar C_{12}^2-\bar C_{22}^2}{\bar C_{11}-\bar C_{22}}\,,\nn\\
\mu_2&=&-\dot\alpha^2-\bar B\dot\alpha-\frac{(\bar C_{11}-\bar C_{22})^2+4C_{12}^2}{\bar C_{11}-\bar C_{22}}\,\cos^2\alpha
+\frac{\bar C_{11}^2-\bar C_{11}\bar C_{22}+2\bar C_{12}^2}{\bar C_{11}-\bar C_{22}}\,.
\end{eqnarray}
It is straightforward to obtain the energy function (which is equal in value to the Hamiltonian, see e.g.~\cite{Goldstein}  for details) which reduces to a formally simple form (see Appendix  \ref{app:H}), namely
\begin{equation}
\label{Hamiltonian}
H(\Phi_i,\dot\Phi_i)=\frac{a^3}2\left[\dot{\Phi}_1^2 +\dot{\Phi}_2^2 +\mu_1(t,k)\,\Phi_1^2 +\mu_2(t,k)\,\Phi_2^2\right]\,,
\end{equation}
so that the Hamiltonian will be unbounded from below if the eigenvalues satisfy $\mu_1<0$ or $\mu_2<0$, for example on the line ($\dot\Phi_i=0$). For $k=0$, we will have a mass instability if $\mu_i(t,0)<0$, and it rapidly evolves  if $|\mu_i(t,0)|\gg H^2$ (strong tachyonic instability). We will then consider unviable those theories which have a strong tachyonic instability, i.e.\ those which possess $\mu_i(t,0)<0$ for which $-\mu_i(t,0)\gg H^2$. However, let us note that the requirement $\mu_i>0$ is too stringent to ensure the short-time stability, indeed one can relax this assumption by requiring that a viable theory, should have eigenvalues (if they are negative) which need to satisfy the condition $|\mu_i(t,0)|\lesssim H^2$, so that the evolution of this instability will not affect the whole stability of the system for time-intervals much shorter than the Hubble time (see also ref.~\cite{Gumrukcuoglu:2016jbh}). 
Additionally, one would expect that the $\mu_2$ eigenvalue is negative as the dust sector will exhibit a Jeans instability. This is necessary in order to guarantee structure formation in our Universe.

\begin{itemize}
\item \textbf{Minimally coupled quintessence model in  presence of a dust fluid}
\end{itemize}

We will now proceed to exemplify the previous, rather abstract, approach by studying a specific model in the presence of dust: minimally coupled quintessence, which has the following action~\cite{Tsujikawa:2013fta}
\begin{align}
S_\phi = \int{}d^4x\sqrt{-g}\l[\frac{m_0^2}{2}R-\f{1}{2}g^{\mu\nu}\partial_\mu\phi\partial_\nu\phi-V(\phi)\r] +S_m\,,
\end{align}
where $\phi$ is the scalar field and $V$ the corresponding potential.
The above can be mapped in the EFT formalism by making the following correspondence~\cite{Gubitosi:2012hu,Bloomfield:2012ff,Gleyzes:2014rba,Frusciante:2016xoj}
\be
c= \frac{1}{2 } \dot{\phi}_0^2\,, \qquad \Lambda=\frac{1}{2} \dot{\phi}_0^2 -V\left(\phi_0\right)\,, \qquad \l\{\Omega,\hat{M}^2,\bar{M}^2_2,\bar{M}^2_3,M^3_1,M^4_2\r\}=0,
\ee 
where $\phi_0(t)$ is the background value of the scalar field. 

Let us now consider that the minimally coupled quintessence model can be also parametrized by assuming that the modification to the gravity sector can be recast as a DE perfect fluid by introducing the following:
\be
w_{\rm DE}(a)\equiv\f{P_{\rm DE}}{\bar{\rho}_{\rm DE}}=\f{\dot{\phi}^2/2-V(\phi)}{\dot{\phi}^2/2+V(\phi)}\,,
\ee
and assuming that the DE density has the standard perfect fluid form
\be
\bar{\rho}_{\rm DE}= 3m_0^2 H_0^2\, \Omega_{\rm DE}^0 \, \exp \bigg[-3 \int_{1}^{a}\frac{(1+w_{\rm DE}(a))}{a} \,da \bigg]\,,
\ee
with $H_0, \Omega_{DE}^0$ be the present day values of the Hubble and density parameter respectively. Then, the Friedmann equation simply reads:
\be
3m_0^2H^2=\bar{\rho}_d+\bar{\rho}_{\rm DE}.
\ee
and the EFT functions can be written accordingly as
\be \label{EFTwDE}
c= \frac{1}{2}\bar{\rho}_{\rm DE}(1+w_{\rm DE})\,, \qquad \Lambda= w_{\rm DE}\bar{\rho}_{\rm DE}.
\ee
This choice for the parametrization makes the whole treatment of the minimally coupled quintessence case more handy. Indeed, this will allow us to rewrite the mass eigenvalues  \eqref{eigenvaluesbhorndeski}, presented in appendix \ref{eigenvalues}, purely in terms of the fluid parameter, i.e.\ $\mu_i(w_{\rm DE})$.
As a general remark, from the expressions \eqref{eigenvaluesbhorndeski} it becomes clear that in general the mass eigenvalues tend to be quite complicated. More complicated theories, especially the non minimally coupled ones, will be substantially harder to treat, yet not impossible. Having the explicit results of the mass eigenvalues for the minimally coupled quintessence model,  we want now to proceed and gain some intuition regarding their behaviour compared to $H^2$.

\begin{figure}[t!]
\begin{center}
  \includegraphics[width=0.45\textwidth]{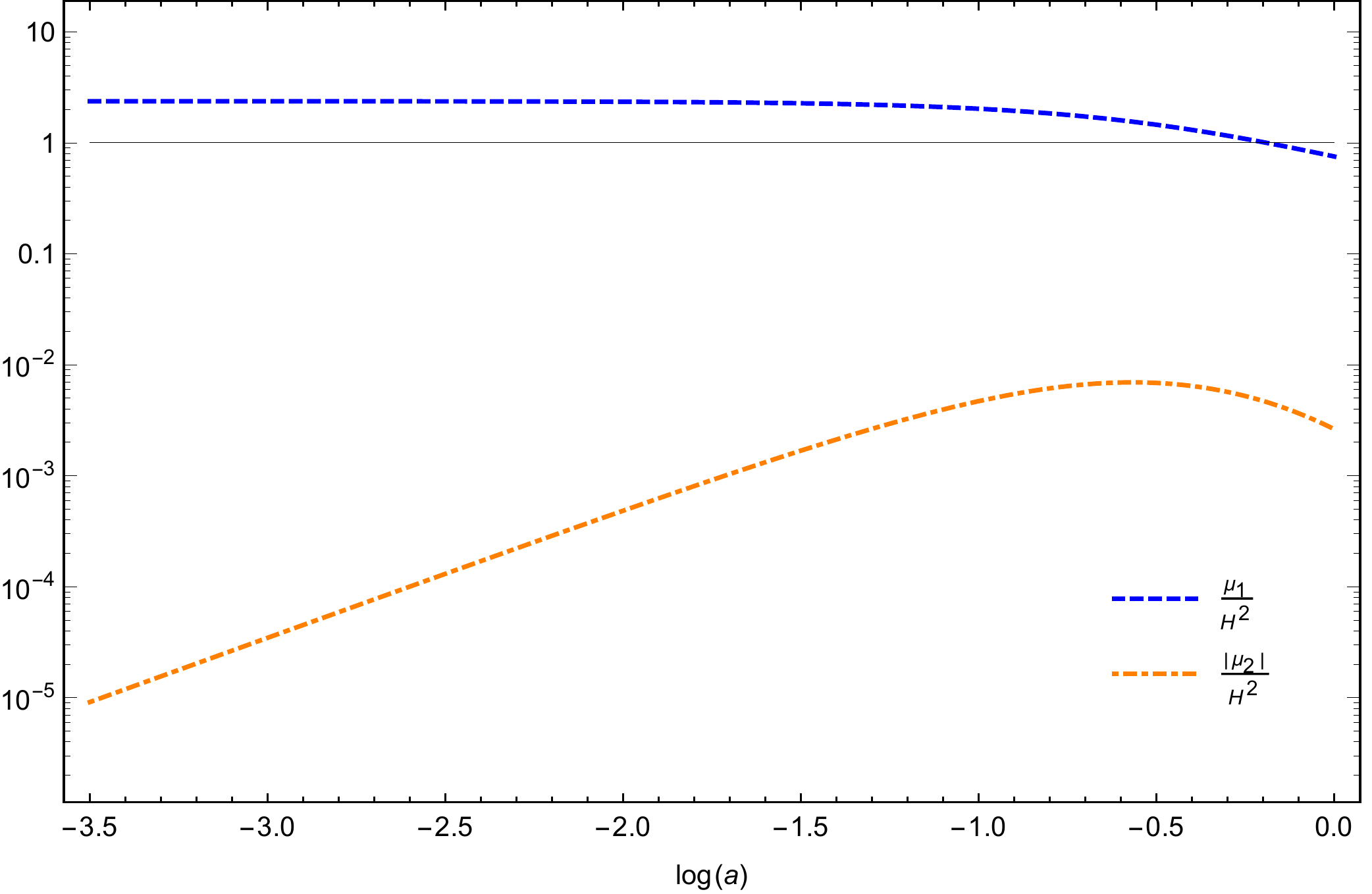}
\includegraphics[width=0.45\textwidth]{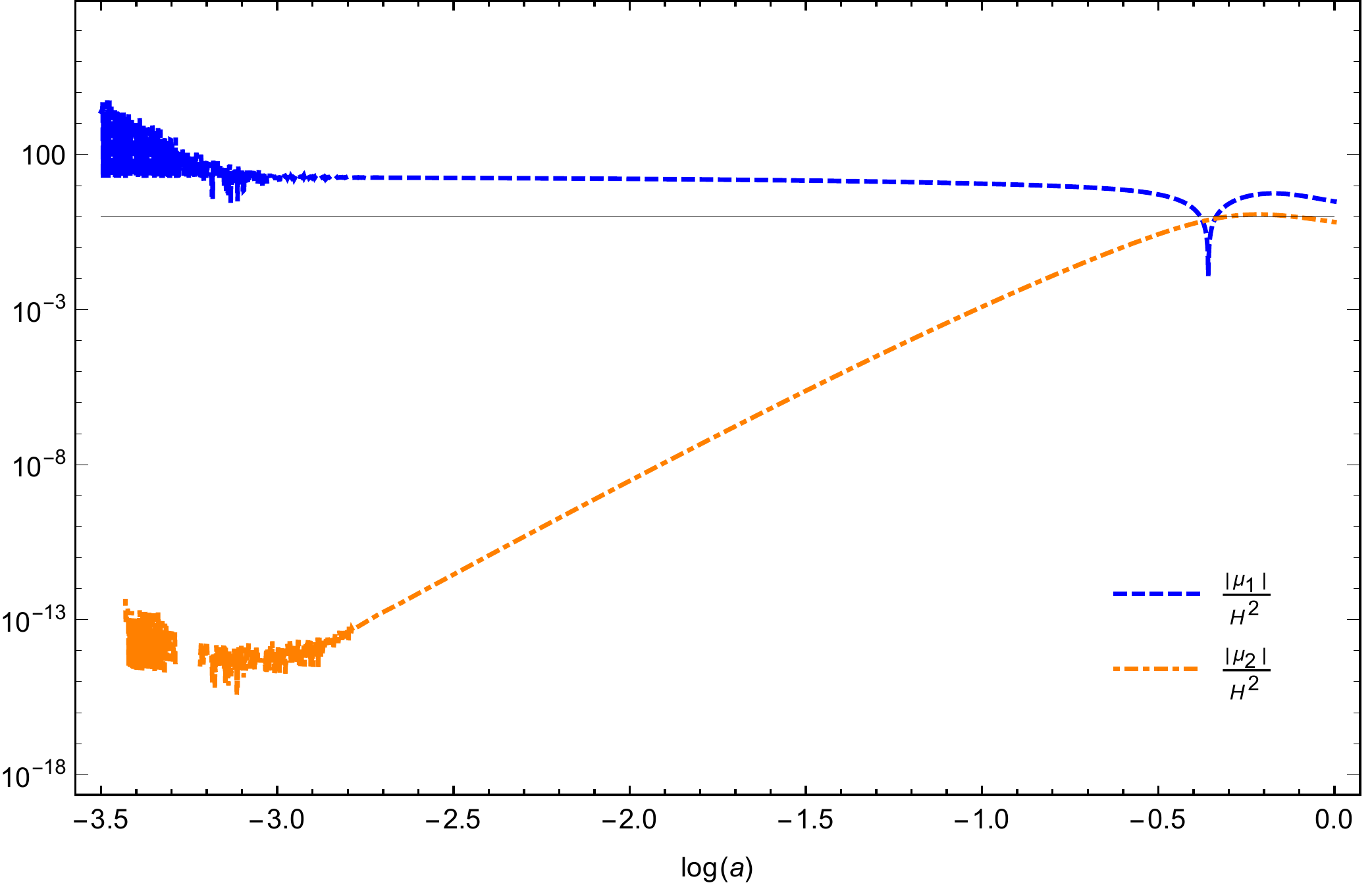}
  \caption{\label{Fig.CPL} The figures show the behaviours of the mass eigenvalues $\mu_1/H^2$ (blue dashed line), $\mu_2/H^2$ (orange dot dashed line) for minimally coupled quintessence on a CPL background. \textit{Left panel}: stable tachyonic configuration with $w_0=-0.9$ and $w_a=0.009$. \textit{Right panel}: unstable tachyonic configuration with $w_0=-2.9$ and $w_a=-2$. For this figure the cosmological parameters are chosen to be: $\Omega_{\rm DE}^0=0.69, \Omega_d^0=0.31, H_0=67.74 $~\cite{Ade:2015xua}.  See section~\ref{Sec:Mass} for the whole discussion.  } 
\end{center}
\end{figure}

We will make a specific choice of the DE equation of state, out of the many options, which will help in illustrating different behaviours: 
\begin{itemize}
\item The CPL parametrization~\cite{Chevallier:2000qy,Linder:2002et}: $w_{\rm DE}(a)=w_0 + w_a(1-a)$, where  $w_0$ and $w_a$ are constant and indicate, respectively, the value and the derivative of $w_{\rm DE}$ today;
\end{itemize}
as illustrative examples, for the values of $\{w_0, w_a\}$ in the DE equation of state we choose two sets, one for which the system  is free from  tachyonic instability and one where the gravity sector shows an unstable configuration since a tachyonic instability is manifest. 
The results are illustrated in figure~\ref{Fig.CPL}. In the left panel, for the choice of the parameters $w_0=-0.9$ and $w_a=0.009$, we  notice that the eigenvalue associated to the gravity sector (i.e.\ $\mu_1$) is always positive and approximately of the same order as $H^2$. On the contrary, the eigenvalue associated with the dust sector ($\mu_2$), here plotted in its absolute value, is negative and $|\mu_2|\ll H^2$. This, of course, has to be expected as it is a manifestation of the well known Jeans instability which allows structure to form. 
 In the right panel, we chose a rather unrealistic set of the parameters in order to show a tachyonic instability, namely $w_0=-2.9$ and $w_a=-2$. Both the eigenvalues oscillate and switch the sign very fast at early time, then $\mu_2$ becomes positive around $\mbox{log}(a)>-2.9$ and $\mu_1$ turns to be firstly negative and finally positive at very late time. In this case since the eigenvalue associated to  gravity is negative (for most of the time) and additionally $|\mu_1|\gg H^2$, this implies that the tachyonic instability evolves very fast, resulting in an unstable system. The dust eigenvalue on the other hand is positive during the matter dominated era which means that matter does not cluster.

The above discussion concerns only the tachyonic and Jeans instabilities, thus can not be considered exhaustive. In order to complete the set of stability conditions for minimally coupled quintessence one needs to study the ghost conditions and the speeds of propagation, as presented in the previous sections,  which in the case of minimally coupled quintessence simply reduce to $w_{\rm DE}(a)>-1$.

\section{Conclusion}\label{Sec:conclusion}

In this paper we have presented a thorough analysis of the viability conditions which  guarantee the stability of the scalar DoFs in the presence of matter fields. As usual, this includes  the avoidance of ghosts and  tachyonic instabilities, supplemented with a positive speed of propagation.
The study of the viability of specific gravity theories in vacuum or in the presence of matter fields has already yielded an extensive literature.  However, our results are more general and  directly applicable to most of the well known models which are of cosmological interest.

 For the gravity sector, we employed the general  EFT approach for DE/MG, which has the advantage of being a model independent parametrization of gravity theories with one extra scalar DoF while at the same time preserving a direct link with a wide class of theoretical models which can be explicitly mapped into this formalism. In order to describe the standard perfect fluids  we chose the Sorkin-Schutz action which has been shown to be well behaved in contrast to other choices made in the past, such as $P(\mathcal{X})$.  In detail, we specialised to the case where the matter fluids are dust (or CDM) and radiation. From these starting blocks we constructed the  Lagrangian and then we proceeded to derive  the action up to second order in scalar perturbations accompanied by the background equations. Finally, we moved to the study of the viability requirements which we will summarise and discuss in the following. 
	
After constructing the Lagrangian for the perturbations one can straightforwardly guarantee the absence of ghosts  by imposing the positivity of the kinetic term, or matrix in case more than one field is considered  as in the present paper. In deriving such conditions we have considered only the Lagrangian in the high-$k$ regime, following the recent results in ref.~\cite{Gumrukcuoglu:2016jbh}. Indeed, it has been shown that only  the high energy terms can turn out in catastrophic instabilities while the sub-leading terms can be recast in mass-like terms through appropriate field redefinitions.  Because the EFT approach encompasses a variety of DE/MG models, which in some cases show different and non trivial k-dependence, it is not possible to obtain one general result applicable to all possible theories. Therefore, we have identified five relevant sub-cases for which we have worked out the corresponding no-ghost conditions. In particular, two of the aforementioned sub-cases correspond to well known theoretical models, i.e.\ low-energy Ho\v rava gravity and beyond Horndeski, while the remaining three do not correspond to any specific class of theories but can be useful in a model independent study. In general, we found three no-ghost conditions for each of the sub-cases,  two of them contain only  matter functions and thus resulting to be trivially satisfied, while the other is more involved as it is a combination of EFT functions.  Finally, we have also identified conditions which lead to strong coupling regimes, thus excluding these theories from an effective description.

The next step was to study the speeds of propagation of the three DoFs, in the high-$k$ limit,  for the  sub-cases  mentioned before which we demanded to be positive.  Depending on the sub-case the results change drastically because the momentum dependence of various terms differs in each sub-case. In general, the gravity speed of propagation  does not depend on fluid variables once one consider theories with higher (than second) order spatial derivatives. In particular, for the sub-case to which low-energy Ho\v rava gravity  belongs, we find that at high-$k$ the DoFs are completely decoupled and the speeds of propagation of dust and radiation components are unaffected by the coupling to gravity, leading to the standard results. While, in the sub-case corresponding to beyond Horndeski only the dust speed of propagation stays unaltered, while both the radiation and gravity speeds are strongly altered due to their interaction. Moreover, when specifying the beyond Horndeski sub-case to the Horndeski one the three DoFs decouple and both dust and radiation show the standard results for speed while the speed of propagation associated to the gravity mode still manifest modification due to the matter components. 
The three remaining sub-cases exhibit unaltered matter speeds and the gravity one is not influenced by the matter sector.  A surprising result is the sub-case corresponding to  $F_4=0, m_2^2\neq0$, for  which  the gravitational sector has a vanishing speed of sound.

In the last part of the work, we have considered the instabilities that show up in case the Hamiltonian is unbounded from below at low energy.  This instability is tightly related with the appearance of a tachyonic mass in the Lagrangian.  Only for this case we have simplified the approach by assuming only one matter fluid, which we chose to be dust.  Consequently, we have identified the two eigenvalues ($\mu_i$) of the system which need to be  constrained in the limit $k\rightarrow 0$ in order to guarantee the boundedness of the Hamiltonian. A stringent condition is to demand both the eigenvalues to be positive definite. On the other hand, it is well known that at early times the dust fluid exhibits a Jeans instability, which is necessary in order to allow  structures formation. Therefore, it is more realistic to assume that the eigenvalues are of the same order of $H^2$, and in case $\mu_i<0$ impose that their evolution is such that $|\mu_i|\ll H^2$. Due to the complexity of these results, we have chosen to exemplify our findings by studying the minimally coupled quintessence case. We have parametrized the gravity modification in terms of the equation of state for DE, i.e.\ $w_{\rm DE}(a)$ and then through the appropriate  mapping, we were able to write $\mu_i(w_{\rm DE})$. Finally, as illustrative example we chose the CPL parametrization for the DE fluid and two sets of values for the DE parameters. In figure~\ref{Fig.CPL}, we showed two typical situations, i.e.\  the case in which the tachyonic instability shows up and the theory becomes pathological and a stable case exhibiting a Jeans instability for the dust sector.

Before concluding, we would like to stress that, although in literature the conditions for no-ghost and a positive speed of propagation are usually considered and deeply studied in specific class of theories, in the present work we showed general results for any theory of gravity with one extra scalar DoF in presence of matter fields. This has been done with the EFT of DE/MG approach and an appropriate choice of the matter DoFs (see section \ref{Sec:matter}). Our choice for the matter DoF allowed us to construct the no-ghost conditions when considering a dust fluid, a result which was not present in the literature before. Moreover, we have also presented the tachyonic conditions which are usually not considered and provide an additional way to explore the parameter space of scalar tensor theories. In some cases, when we specialize the EFT functions to specific sub-cases, we found our results to be compatible with other results already present in literature.  In those cases the main improvement coming from our approach is the fact we work directly with the density perturbation as the fluid variable, hence avoiding issues with the definition of the matter Lagrangian and, subsequently, any need to take limits in order to include  pressureless fluids. The lack of such ambiguities solidifies the pre-existing  results and guarantees their accuracy.

As a final remark, we would like to stress that, due to the generality of the approach we used, one can safely apply our results to most of the well known cosmological models {when any matter species is present, once the mapping between the chosen model and the EFT functions is known. Therefore, they can be employed  by cosmological tools, such as EFTCAMB, in order to reduce the viability space and ensure that the model under consideration is free of any pathological instabilities.

\begin{acknowledgments}
We are grateful to Bin Hu, Shinji Mukohyama for useful discussions. We are in debt to Alessandra Silvestri for her contribution at early stage of the work and we acknowledge her for comments on the manuscript. ADF is grateful to the Institute for Fundamental Study of Naresuan University for warm hospitality where this work was advanced.
ADF was supported by JSPS KAKENHI Grant Numbers 16K05348, 16H01099. The research of NF has received funding from the European Research Council under the European Community's Seventh Framework Programme (FP7/2007-2013, Grant Agreement No.~307934). GP acknowledges support from the D-ITP consortium, a program of the Netherlands Organisation for Scientific Research (NWO) that is funded by the Dutch Ministry of Education, Culture and Science (OCW). NF and GP acknowledge the COST Action (CANTATA/CA15117), supported by COST (European Cooperation in Science and Technology).
\end{acknowledgments}

\appendix  
\section{Matrix coefficients }\label{APP:Coefficients}

For completeness in this Appendix we will explicitly list the matrix coefficients used in sections \ref{Sec:stability},\ref{Sec:speed} and \ref{Sec:Mass}.
Let us start by considering  the action (\ref{actionshort}) introduced in section \ref{Sec:stability}:
\begin{equation}
S^{(2)}=\f{1}{(2\pi)^3}\int{} d^3kdta^3\l(\dot{\vec{\chi}}^t\textbf{A}
\dot{\vec{\chi}}-k^2\vec{\chi}^t\textbf{G}\vec{\chi}-
\dot{\vec{\chi}}^t\textbf{B}\vec{\chi}-\vec{\chi}^t\textbf{M}\vec{\chi}\r)\,,
\end{equation}
where the coefficients are
\ba
&&A_{11}=\frac{3 \left(F_1-3 F_{4}\right) \left(k^2 a^2 \left(-8 m_{2}^2 \left(3 \bar{\rho}_d+4 \bar{\rho} _r\right)+3 F_2^2+F_1 F_3\right)-a^4 F_3 \left(3 \bar{\rho} _d+4 \bar{\rho} _r\right)+8 k^4 F_1 m_{2}^2\right)}{2 k^2 a^2 \left(-8 m_{2}^2 \left(3 \bar{\rho} _d+4 \bar{\rho} _r\right)+3 F_3 F_{4}+3 F_2{}^2\right)-2 a^4 F_3 \left(3 \bar{\rho} _d+4 \bar{\rho} _r\right)+48 k^4 m_{2}^2 F_{4}}\,, \nn\\
&&A_{12}=A_{21}=-\frac{3 a^2 \bar{\rho} _d \left(F_1-3 F_{4}\right) \left(a^2 F_3+8 k^2 m_{2}^2\right)}{2 k^2 a^2 \left(-8 m_{2}^2 \left(3 \bar{\rho} _d+4 \bar{\rho} _r\right)+3 F_3 F_{4}+3 F_2^2\right)-2 a^4 F_3 \left(3 \bar{\rho}_d+4 \bar{\rho} _r\right)+48 k^4 m_{2}^2 F_{4}}\,, \nn\\
&&A_{13}=A_{31}=-\frac{3 a^2 \left(F_1-3 F_{4}\right) \bar{\rho} _r \left(a^2 F_3+8 k^2 m_{2}^2\right)}{2 k^2 a^2 \left(-8 m_{2}^2 \left(3 \bar{\rho}_d+4 \bar{\rho}_r\right)+3 F_3 F_{4}+3 F_2^2\right)-2 a^4 F_3 \left(3 \bar{\rho}_d+4 \bar{\rho}_r\right)+48 k^4 m_{2}^2 F_{4}} \,,\nn\\
&&A_{22}=\frac{a^2 \bar{\rho} _d \left(k^2 a^2 \left(3 F_3 F_{4}+3 F_2^2-32 m_{2}^2 \bar{\rho}_r\right)-4 a^4 F_3 \bar{\rho}_r+24 k^4 m_{2}^2 F_{4}\right)}{2 k^2 \left(k^2 a^2 \left(-8 m_{2}^2 \left(3 \bar{\rho}_d+4 \bar{\rho}_r\right)+3 F_3 F_{4}+3 F_2^2\right)-a^4 F_3 \left(3 \bar{\rho}_d+4 \bar{\rho}_r\right)+24 k^4 m_{2}^2 F_{4}\right)} \,,\nn\\
&&A_{23}=A_{32}=\frac{3 a^4 \bar{\rho}_d \bar{\rho}_r \left(a^2 F_3+8 k^2 m_{2}^2\right)}{2 k^2 \left(k^2 a^2 \left(-8 m_{2}^2 \left(3 \bar{\rho}_d+4 \bar{\rho}_r\right)+3 F_3 F_{4}+3 F_2^2\right)-a^4 F_3 \left(3 \bar{\rho}_d+4 \bar{\rho}_r\right)+24 k^4 m_{2}^2 F_{4}\right)}\,, \nn\\
&&A_{33}= \frac{9 a^2 \bar{\rho}_r \left(k^2 a^2 \left(-8 m_{2}^2 \bar{\rho}_d+F_3 F_{4}+F_2^2\right)-a^4 F_3 \bar{\rho}_d+8 k^4 m_{2}^2 F_{4}\right)}{8 k^2 \left(k^2 a^2 \left(-8 m_{2}^2 \left(3 \bar{\rho}_d+4 \bar{\rho} _r\right)+3 F_3 F_{4}+3 F_2^2\right)-a^4 F_3 \left(3 \bar{\rho}_d+4 \bar{\rho}_r\right)+24 k^4 m_{2}^2 F_{4}\right)}\,,\nn\\
&&B_{11}=-\frac{6 k^4 F_2 \left(F_1-3 F_{4}\right) \left(2 \hat{M}^2+m_0^2 (\Omega +1)\right)}{k^2 a^2 \left(-8 m_2^2 \left(3 \bar{\rho} _d+4 \bar{\rho} _r\right)+3 F_3 F_{4}+3 F_2{}^2\right)-a^4 F_3 \left(3 \bar{\rho} _d+4 \bar{\rho} _r\right)+24 k^4 m_2^2 F_{4}}\,, \nn\\
&&B_{12}= \frac{3 k^2 a^2 F_2 \bar{\rho} _d \left(F_1-3 F_{4}\right)}{k^2 a^2 \left(-8 m_2^2 \left(3 \bar{\rho} _d+4 \bar{\rho} _r\right)+3 F_3 F_{4}+3 F_2{}^2\right)-a^4 F_3 \left(3 \bar{\rho} _d+4 \bar{\rho} _r\right)+24 k^4 m_2^2 F_{4}}\,,\nn\\
&&B_{13}= \frac{3 k^2 a^2 F_2 \left(F_1-3 F_{4}\right) \bar{\rho} _r}{k^2 a^2 \left(-8 m_2^2 \left(3 \bar{\rho} _d+4 \bar{\rho} _r\right)+3 F_3 F_{4}+3 F_2{}^2\right)-a^4 F_3 \left(3 \bar{\rho} _d+4 \bar{\rho} _r\right)+24 k^4 m_2^2 F_{4}}\,,\nn\\
&&B_{22}= \frac{3 a^4 F_2 \bar{\rho} _d{}^2}{k^2 a^2 \left(8 m_2^2 \left(3 \bar{\rho} _d+4 \bar{\rho} _r\right)-3 F_3 F_{4}-3 F_2{}^2\right)+a^4 F_3 \left(3 \bar{\rho} _d+4 \bar{\rho} _r\right)-24 k^4 m_2^2 F_{4}}\,,\nn\\
&&B_{21}=\frac{6 k^2 a^2 F_2 \bar{\rho} _d \left(2 \hat{M}^2+m_0^2 (\Omega +1)\right)}{k^2 a^2 \left(-8 m_2^2 \left(3 \bar{\rho} _d+4 \bar{\rho} _r\right)+3 F_3 F_{4}+3 F_2{}^2\right)-a^4 F_3 \left(3 \bar{\rho} _d+4 \bar{\rho} _r\right)+24 k^4 m_2^2 F_{4}}\,,\nn\\
&&B_{23}=B_{32}= \frac{3 a^4 F_2 \bar{\rho} _d \bar{\rho} _r}{k^2 a^2 \left(8 m_2^2 \left(3 \bar{\rho} _d+4 \bar{\rho} _r\right)-3 F_3 F_{4}-3 F_2{}^2\right)+a^4 F_3 \left(3 \bar{\rho} _d+4 \bar{\rho} _r\right)-24 k^4 m_2^2 F_{4}}\,,\nn\\
&&B_{33}= \frac{3 a^4 F_2 \bar{\rho} _r{}^2}{k^2 a^2 \left(8 m_2^2 \left(3 \bar{\rho} _d+4 \bar{\rho} _r\right)-3 F_3 F_{4}-3 F_2{}^2\right)+a^4 F_3 \left(3 \bar{\rho} _d+4 \bar{\rho} _r\right)-24 k^4 m_2^2 F_{4}}\,,\nn\\
&&B_{31}=\frac{6 k^2 a^2 F_2 \bar{\rho} _r \left(2 \hat{M}^2+m_0^2 (\Omega +1)\right)}{k^2 a^2 \left(-8 m_2^2 \left(3 \bar{\rho} _d+4 \bar{\rho} _r\right)+3 F_3 F_{4}+3 F_2^2\right)-a^4 F_3 \left(3 \bar{\rho} _d+4 \bar{\rho} _r\right)+24 k^4 m_2^2 F_{4}}\,,\nn\\
&&G_{11}=\l\{k^2 a^2 \left(m_0^2 (\Omega +1) \left(-8 \left(3 \bar{\rho} _d+4 \bar{\rho} _r\right) \left(m_2^2-\hat{M}^2\right)+3 F_3 F_{4}+3 F_2{}^2\right)+8 \hat{M}^4 \left(3 \bar{\rho} _d+4 \bar{\rho} _r\right)+2 m_0^4 (\Omega +1)^2 \left(3 \bar{\rho} _d+4 \bar{\rho} _r\right)\right)\r.\nn\\
&&\l.-m_0^2 a^4 F_3 (\Omega +1) \left(3 \bar{\rho} _d+4 \bar{\rho} _r\right)-6 k^4 F_{4} \left(-4 m_0^2 (\Omega +1) \left(m_2^2-\hat{M}^2\right)+4 \hat{M}^2{}^2+m_0^4 (\Omega +1)^2\right)\r\}/\l\{a^2 \left(k^2 a^2 \left(8 m_2^2 \left(3 \bar{\rho} _d+4 \bar{\rho} _r\right)\r.\r.\r.\nn\\
&&\l.\l.\l.-3 F_3 F_{4}-3 F_2{}^2\right)+a^4 F_3 \left(3 \bar{\rho} _d+4 \bar{\rho} _r\right)-24 k^4 m_2^2 F_{4}\right)\r\}\,,\nn\\
&&G_{12}=G_{21}= -\frac{\bar{\rho} _d \left(2 \hat{M}^2+m_0^2 (\Omega +1)\right) \left(-a^2 \left(3 \bar{\rho} _d+4 \bar{\rho} _r\right)+3 k^2 \bar{M}_2^2+3 k^2 \bar{M}_3^2\right)}{k^2 a^2 \left(-8 m_2^2 \left(3 \bar{\rho} _d+4 \bar{\rho} _r\right)+3 F_3 F_{4}+3 F_2{}^2\right)-a^4 F_3 \left(3 \bar{\rho} _d+4 \bar{\rho} _r\right)+24 k^4 m_2^2 F_{4}}\,,\nn\\
&&G_{13}= G_{31}=-\frac{\bar{\rho} _r \left(2 \hat{M}^2+m_0^2 (\Omega +1)\right) \left(-a^2 \left(3 \bar{\rho} _d+4 \bar{\rho} _r\right)+3 k^2 \bar{M}_2^2+3 k^2 \bar{M}_3^2\right)}{k^2 a^2 \left(-8 m_2^2 \left(3 \bar{\rho} _d+4 \bar{\rho} _r\right)+3 F_3 F_{4}+3 F_2{}^2\right)-a^4 F_3 \left(3 \bar{\rho} _d+4 \bar{\rho} _r\right)+24 k^4 m_2^2 F_{4}}\,,\nn\\
&&G_{22}= \frac{3 a^2 F_{4} \bar{\rho} _d{}^2}{2 k^2 a^2 \left(-8 m_2^2 \left(3 \bar{\rho} _d+4 \bar{\rho} _r\right)+3 F_3 F_{4}+3 F_2{}^2\right)-2 a^4 F_3 \left(3 \bar{\rho} _d+4 \bar{\rho} _r\right)+48 k^4 m_2^2 F_{4}}\,,\nn\\
&&G_{23}= G_{32}=\frac{3 a^2 F_{4} \bar{\rho} _d \bar{\rho} _r}{2 k^2 a^2 \left(-8 m_2^2 \left(3 \bar{\rho} _d+4 \bar{\rho} _r\right)+3 F_3 F_{4}+3 F_2{}^2\right)-2 a^4 F_3 \left(3 \bar{\rho} _d+4 \bar{\rho} _r\right)+48 k^4 m_2^2 F_{4}}\,,\nn\\
&&G_{33}= \frac{\bar{\rho} _r \left(a^2 \left(-4 \left(m_2^2 \left(6 \bar{\rho} _d+8 \bar{\rho} _r\right)-3 F_{4} \bar{\rho} _r\right)+3 F_3 F_{4}+3 F_2{}^2\right)+24 k^2 m_2^2 F_{4}\right)}{8 \left(k^2 a^2 \left(-8 m_2^2 \left(3 \bar{\rho} _d+4 \bar{\rho} _r\right)+3 F_3 F_{4}+3 F_2{}^2\right)-a^4 F_3 \left(3 \bar{\rho} _d+4 \bar{\rho} _r\right)+24 k^4 m_2^2 F_{4}\right)}\,,\nn\\
&&M_{11}=M_{12}=M_{21}=M_{13}=M_{31}=0\,,\nn\\
&&M_{22}=-\frac{a^4 \bar{\rho} _d{}^2 \left(3 \bar{\rho} _d+4 \bar{\rho} _r\right)}{2 k^2 a^2 \left(-8 m_2^2 \left(3 \bar{\rho} _d+4 \bar{\rho} _r\right)+3 F_3 F_{4}+3 F_2{}^2\right)-2 a^4 F_3 \left(3 \bar{\rho} _d+4 \bar{\rho} _r\right)+48 k^4 m_2^2 F_{4}}\,,\nn\\
&&M_{23}=M_{32}=-\frac{a^4 \bar{\rho} _d \bar{\rho} _r \left(3 \bar{\rho} _d+4 \bar{\rho} _r\right)}{2 k^2 a^2 \left(-8 m_2^2 \left(3 \bar{\rho} _d+4 \bar{\rho} _r\right)+3 F_3 F_{4}+3 F_2{}^2\right)-2 a^4 F_3 \left(3 \bar{\rho} _d+4 \bar{\rho} _r\right)+48 k^4 m_2^2 F_{4}}\,,\nn\\
&&M_{33}=-\frac{a^4 \bar{\rho} _r \left(3 \bar{\rho} _d+4 \bar{\rho} _r\right) \left(F_3+4 \bar{\rho} _r\right)}{8 \left(k^2 a^2 \left(-8 m_2^2 \left(3 \bar{\rho} _d+4 \bar{\rho} _r\right)+3 F_3 F_{4}+3 F_2{}^2\right)-a^4 F_3 \left(3 \bar{\rho} _d+4 \bar{\rho} _r\right)+24 k^4 m_2^2 F_{4}\right)}.
\ea

Now, we write down the matrix coefficients of eq.  (\ref{actionmass}) in section \ref{Sec:Mass}:
\begin{equation}
  \mathcal{L}^{(2)}=\frac{a^3}2\left[\dot{\bar\Psi}_1^2 +\dot{\bar\Psi}_2^2 + \bar B(t,k)\,(\dot{\bar\Psi}_1{\bar\Psi}_2-\dot{\bar\Psi}_2{\bar\Psi}_1)-\bar C_{ij}(t,k){\bar\Psi}_i{\bar\Psi}_j\right],
\end{equation}
where 
\ba
&&\bar{B}=-\frac{k \left(A_{22} \left(-2 \dot{A}_{12}+B_{12}-B_{21}\right)+2 A_{12} \dot{A}_{22}\right)}{4 A_{22} \sqrt{\bar{K}_{11}} \sqrt{\bar{K}_{22}}} \,,\nn\\
&&\bar{C}_{12}=\bar{C}_{21}=\{k \left(a \left(A_{22} \left(A_{12} \left(\bar{K}_{11} \left(4 \bar{K}_{22} \left(\ddot{A}_{22}+\dot{B}_{22}-2 M_{22}\right)+2 \dot{A}_{22} \dot{\bar{K}}_{22}\right)-2 \bar{K}_{22} \dot{A}_{22} \dot{\bar{K}}_{11}-8 k^2 G_{22} \bar{K}_{11} \bar{K}_{22}\right)\r.\r.\r.\nn\\
&&\l.\l.\l.+4 \bar{K}_{11} \bar{K}_{22} \dot{A}_{12} \dot{A}_{22}\right)+A_{22}^2 \left(-\bar{K}_{11} \left(2 \bar{K}_{22} \left(2 \ddot{A}_{12}+\dot{B}_{12}+\dot{B}_{21}\right)+\dot{\bar{K}}_{22} \left(2 \dot{A}_{12}-B_{12}+B_{21}\right)\right)\r.\r.\r.\nn\\
&&\l.\l.\l.+\bar{K}_{22} \dot{\bar{K}}_{11} \left(2 \dot{A}_{12}-B_{12}+B_{21}\right)+8 k^2 G_{12} \bar{K}_{11} \bar{K}_{22}\right)-4 A_{12} \bar{K}_{11} \bar{K}_{22} \dot{A}_{22}^2\right)-6 A_{22} \bar{K}_{11} \bar{K}_{22} aH \left(A_{22} \left(2 \dot{A}_{12}+B_{12}+B_{21}\right)\r.\r.\nn\\
&&\l.\l.-2 A_{12} \left(\dot{A}_{22}+B_{22}\right)\right)\right)\}/\{16 a A_{22}^2 \bar{K}_{11}^{3/2} \bar{K}_{22}^{3/2}\}\,,\nn\\
&&\bar{C}_{11}=\l[6 A_{22} \bar{K}_{11} H \left(A_{22} A_{12} \left(A_{12} \dot{\bar{K}}_{11}+B_{12} \bar{K}_{11}+B_{21} \bar{K}_{11}\right)-A_{22}{}^2 \left(A_{11} \dot{\bar{K}}_{11}+B_{11} \bar{K}_{11}\right)-A_{12}{}^2 B_{22} \bar{K}_{11}\right) \r.\nn\\
&&\l.+ 2 A_{12} A_{22} \bar{K}_{11} \left(A_{12} \left(-\dot{A}_{22} \dot{\bar{K}}_{11}+\bar{K}_{11} \left(2 M_{22}-\dot{B}_{22}\right)+2 k^2 G_{22} \bar{K}_{11}\right)+\bar{K}_{11} \dot{A}_{22} \left(4 \dot{A}_{12}-B_{12}+B_{21}\right)\right) \r.\nn\\
&&\l.+A_{22}{}^3 \left(-2 \bar{K}_{11} \left(\dot{A}_{11} \dot{\bar{K}}_{11}+A_{11} \ddot{\bar{K}}_{11}\right)+3 A_{11} \dot{\bar{K}}_{11}{}^2-2 \bar{K}_{11}{}^2 B_{11}'+4 k^2 G_{11} \bar{K}_{11}{}^2\right)\r.\nn\\
&&\l.+A_{22}{}^2 \left(2 A_{12} \bar{K}_{11} \left(2 \dot{A}_{12} \dot{\bar{K}}_{11}+\bar{K}_{11} \left(\dot{B}_{12}+\dot{B}_{21}\right)-4 k^2 G_{12} \bar{K}_{11}\right)+2 \bar{K}_{11}{}^2 \dot{A}_{12} \left(-2 \dot{A}_{12}+B_{12}-B_{21}\right)\r.\r.\nn\\
&&\l.\l.+A_{12}{}^2 \left(2 \bar{K}_{11} \ddot{\bar{K}}_{11}-3 \dot{\bar{K}}_{11}{}^2\right)\right)-4 A_{12}{}^2 \bar{K}_{11}{}^2 \dot{A}_{22}{}^2\r]/\l\{8  A_{22}{}^3 \bar{K}_{11}{}^3\r\}\,,\nn\\
&&\bar{C}_{22}=\frac{k^2}{8  \bar{K}_{22}{}^3} \left[ -2 \bar{K}_{22} \left(\dot{A}_{22} \dot{\bar{K}}_{11}+A_{22} \ddot{\bar{K}}_{11}\right)+3 A_{22} \dot{\bar{K}}_{11}{}^2+\bar{K}_{22}{}^2 \left(4 M_{22}-2 \dot{B}_{22}\right)+4 k^2 G_{22} \bar{K}_{22}{}^2\r.\nn\\&&\l.-6 \bar{K}_{22} H \left(A_{22} \dot{\bar{K}}_{11}+B_{22} \bar{K}_{22}\right)\right].
\ea
Note that the $A_{ij}, B_{ij},G_{ij},M_{ij}$ matrix components that appear in the last four coefficients have been obtained from the full expressions defined above by setting $\bar{\rho}_r=0$.

\section{Obtaining the Hamiltonian}\label{app:H}

In this appendix, we will present the derivation of the Hamiltonian used in section \ref{Sec:Mass} and explain why the antisymmetric matrix B does not affect the unboundedness of the Hamiltonian. For this purpose we star from the Lagrangian \eqref{finalLagr}:
\begin{equation}
\mathcal{L}^{(2)} = \frac{a^3}2\left[\dot{\Phi}_1^2 +\dot{\Phi}_2^2 + B(t,k)\,(\dot{\Phi}_1{\Phi}_2-\dot{\Phi}_2{\Phi}_1)-\mu_1(t,k)\Phi_1^2
-\mu_2(t,k)\Phi_2^2\right].
\end{equation}
Defining, the canonical momenta as:
\begin{align}
p_1&=a^3\l(\dot{\Phi}_1+B(t,k)\Phi_2\r),\nn\\
p_2&=a^3\l(\dot{\Phi}_2-B(t,k)\Phi_1\r),
\end{align}
the Hamiltonian can be written as follows
\begin{align}
H&=\l\{p_1\l(\f{p_1}{a^3}-B\Phi_2\r)+p_2\l(\f{p_2}{a^3}+B\Phi_1\r)-\f{a^3}{2}\l[\l(\f{p_1}{a^3}-B\Phi_2\r)^2+\l(\f{p_2}{a^3}+B\Phi_1\r)^2\r.\r.\nn\\
&\l.\l.+B\l[\l(\f{p_1}{a^3}-B\Phi_2\r)\Phi_2-\l(\f{p_1}{a^3}+B\Phi_1\r)\Phi_1\r]-\mu_1 \Phi_1^2-\mu_2\Phi_2^2\r]\r\}\nn\\
&=\f{a^3}{2}\l[ \l(\f{p_1}{a^3}-B\Phi_2\r)^2+\l(\f{p_2}{a^3}+B\Phi_1\r)^2+\mu_1\Phi_1^2+\mu_2 \Phi_2^2 \r]\,.
\end{align}
Now, in terms of $\{\dot{\Phi}_i,\Phi_i \}$ the above Hamiltonian becomes \eqref{Hamiltonian}:
\begin{equation}
H(\Phi_i,\dot\Phi_i)=\frac{a^3}2\left[\dot{\Phi}_1^2 +\dot{\Phi}_2^2 +\mu_1(t,k)\,\Phi_1^2 +\mu_2(t,k)\,\Phi_2^2\right].
\end{equation}
From this expression it is clear that the antisymmetric matrix does not influence the unboundedness from below of the Hamiltonian, instead such issues are encoded within the functions $\mu_i$.

\section{Mass eigenvalues for beyond Horndeski case }\label{eigenvalues}

In section \ref{Sec:Mass} we studied the mass eigenvalues of the EFT in the presence of a dust fluid. We have presented the procedure and the results in the  case of minimally coupled quintessence but refrained from showing general expressions due to their complexity. Here we present the mass eigenvalues for the beyond Horndeski theories which, when using the appropriate mapping, will yield the previously discussed quintessence results. Then, the eigenvalues are: 
 \begin{align}\label{eigenvaluesbhorndeski}
\mu_1=&\left(-\frac{4 F_1 F_3 \left(F_2 \left(F_3 \dot{F}_1+F_1 \dot{F}_3\right)-2 F_1 F_3 \dot{F}_2\right){}^2 F_2{}^2}{3 F_2{}^2+F_1 F_3}+\frac{4 F_1F_3 \left(F_2 \left(F_3 \dot{F}_1+F_1 \dot{F}_3\right)-2 F_1 F_3 \dot{F}_2\right){}^2}{\frac{F_1 F_3}{F_2{}^2}+3}+\left( F_1 F_2{}^2\times\right.\right.\nn
\\
&\left.\left.\left(6 F_1 F_3 \left(3 F_2{}^2+F_1 F_3\right) \dot{a} \left(3 \dot{F}_3 F_2{}^2+6 F_3 \left(\bar{\rho} _d-\dot{F}_2\right) F_2-2 F_3{}^2 \dot{F}_1\right) F_2{}^2+a\left(6 F_3{}^3 \dot{F}_1{}^2 F_2{}^4+F_1{}^2 F_3 \left(\left(6 F_3 \ddot{F}_3\right.\right.\right.\right.\right.\right.\nn
\\
&\left.\left.\left.\left.\left.\left.-9 \dot{F}_3{}^2\right) F_2{}^3+12 F_3 \left(\left(\dot{F}_2-\bar{\rho} _d\right) \dot{F}_3+F_3 \left(\dot{\bar{\rho}} _d-\ddot{F}_2\right)\right) F_2{}^2+4 F_3{}^2 \left(3 \bar{\rho} _d \left(\dot{F}_2-\bar{\rho} _d\right)+\dot{F}_1 \dot{F}_3-F_3 \ddot{F}_1\right) F_2\right.\right.\right.\right.\right.\nn
\\
&\left.\left.\left.\left.\left.-8 F_3{}^3 \dot{F}_1 \dot{F}_2\right) F_2+2 F_1{}^3 F_3{}^2 \left(F_2 \dot{F}_3-2 F_3 \dot{F}_2\right){}^2+F_1 \left(9 \left(2 F_3 \ddot{F}_3-3 \dot{F}_3{}^2\right) F_2{}^6+36 F_3 \left(\left(\dot{F}_2-\bar{\rho} _d\right) \dot{F}_3+F_3 \left(\dot{\bar{\rho}} _d\right.\right.\right.\right.\right.\right.\right.\nn
\\
&\left.\left.\left.\left.\left.\left.\left.-\ddot{F}_2\right)\right) F_2{}^5-12 F_3{}^2 \left(3 \bar{\rho} _d \left(\bar{\rho} _d-\dot{F}_2\right)+F_3 \ddot{F}_1\right) F_2{}^4+4 F_3{}^4 \dot{F}_1{}^2 F_2{}^2\right)\right)\right)\right)/\left(a \left(3 F_2{}^2+F_1 F_3\right) \left(3 F_2{}^2+2 F_1 F_3\right)\right)-\right.\nn
\\
&\left.\left(9 F_2{}^6 \left(1+\frac{1}{3 \sqrt{\frac{F_2{}^4}{\left(3 F_2{}^2+2 F_1 F_3\right){}^2}}}\right) \left(6 F_1 F_2 F_3 \left(3 F_2{}^2+F_1 F_3\right) \dot{a} \left(F_2 F_3 \dot{F}_1+F_1 \left(2 F_3 \left(\bar{\rho} _d-\dot{F}_2\right)+F_2 \dot{F}_3\right)\right)+\right.\right.\right.\nn
\\
&\left.\left.\left.a \left(-3 F_3{}^2 \dot{F}_1{}^2 F_2{}^4+2 F_1 F_3{}^2 \left(3 F_2{}^2 \ddot{F}_1-F_3\dot{F}_1{}^2\right) F_2{}^2+F_1{}^2 \left(\left(6 F_3 \ddot{F}_3-9 \dot{F}_3{}^2\right) F_2{}^3+12 F_3 \left(\left(\dot{F}_2-\bar{\rho} _d\right) \dot{F}_3+F_3 \left(\dot{\bar{\rho} _d}\right.\right.\right.\right.\right.\right.\right.\nn
\\
&\left.\left.\left.\left.\left.\left.\left.-\ddot{F}_2\right)\right) F_2{}^2+2 F_3{}^2 \left(6 \bar{\rho} _d \left(\dot{F}_2-\bar{\rho} _d\right)-\dot{F}_1 \dot{F}_3+F_3 \ddot{F}_1\right) F_2+4 F_3{}^3 \dot{F}_1 \dot{F}_2\right) F_2+2 F_1{}^3 F_3 \left(-2 \left(\bar{\rho} _d{}^2-\dot{F}_2 \bar{\rho} _d+\dot{F}_2{}^2\right.\right.\right.\right.\right.\right.\nn
\\
&\left.\left.\left.\left.\left.\left.+F_2 \left(\ddot{F}_2-\dot{\bar{\rho} }_d\right)\right) F_3{}^2+F_2 \left(F_2 \ddot{F}_3-2 \left(\bar{\rho} _d-2 \dot{F}_2\right) \dot{F_3}\right) F_3-2 F_2{}^2 \dot{F}_3{}^2\right)\right)\right)\right)/\left(a \left(3 F_2{}^2+F_1 F_3\right){}^2 \left(3 F_2{}^2\right.\right.\right.\nn
\\
&\left.\left.\left.+2 F_1 F_3\right)\right)\right)\frac{1}{\left(16 F_1{}^2 F_2{}^4 F_3{}^2\right)}\,,\nn
\\
\mu_2&= \left(-\frac{4 F_1 F_3 \left(F_2 \left(F_3 \dot{F}_1+F_1 \dot{F}_3\right)-2 F_1 F_3 \dot{F}_2\right){}^2 F_2{}^2}{3 F_2{}^2+F_1 F_3}+\frac{4 F_1 F_3( \left(F_2 \left(F_3 \dot{F}_1+F_1 \dot{F}_3\right)-2 F_1 F_3 \dot{F}_2\right){}^2}{\frac{F_1 F_3}{F_2{}^2}+3}+\left(9 F_2{}^6 \left(1\right.\right.\right.\nn
\\
&\left.\left.\left.+\frac{1}{3 \sqrt{\frac{F_2{}^4}{\left(3 F_2{}^2+2 F_1 F_3\right){}^2}}}\right) \left(6 F_1 F_2 F_3\left(3 F_2{}^2+F_1 F_3\right) \dot{a} \left(F_2 F_3 \dot{F}_1+F_1 \left(2 F_3 \left(\bar{\rho} _d-\dot{F}_2\right)+F_2 \dot{F}_3\right)\right)+a \left(-3 F_3{}^2 \dot{F}_1{}^2 F_2{}^4\right.\right.\right.\right.\nn
\\
&\left.\left.\left.\left.+2 F_1 F_3{}^2 \left(3 F_2{}^2 \ddot{F}_1-F_3 \dot{F}_1{}^2\right) F_2{}^2+F_1{}^2 \left(\left(6 F_3 \ddot{F}_3-9 \dot{F}_3{}^2\right) F_2{}^3+12 F_3 \left(\left(\dot{F}_2-\bar{\rho} _d\right) \dot{F}_3+F_3 \left(\dot{\bar{\rho}}_d-\ddot{F}_2\right)\right) F_2{}^2\right.\right.\right.\right.\right.\nn
\\
&\left.\left.\left.\left.\left.+2 F_3{}^2 \left(6 \bar{\rho} _d\left(\dot{F}_2-\bar{\rho} _d\right)-\dot{F}_1 \dot{F}_3+F_3 \ddot{F}_1\right) F_2+4 F_3{}^3 \dot{F}_1 \dot{F}_2\right) F_2+2 F_1{}^3 F_3 \left(-2 \left(\bar{\rho} _d{}^2-\dot{F}_2 \bar{\rho} _d+\dot{F}_2{}^2+F_2 \left(\ddot{F}_2\right.\right.\right.\right.\right.\right.\right.\nn
\\
&\left.\left.\left.\left.\left.\left.\left.-\dot{\bar{\rho}} _d\right)\right) F_3{}^2+F_2 \left(F_2 \ddot{F}_3-2 \left(\bar{\rho} _d-2 \dot{F}_2\right) \dot{F}_3\right) F_3-2 F_2{}^2 \dot{F}_3{}^2\right)\right)\right)\right)/\left(a \left(3 F_2{}^2+F_1 F_3\right){}^2 \left(3 F_2{}^2+2 F_1 F_3\right)\right)-\left(2 F_2{}^2 F_3 \times\right.\right.\nn
\\
&\left.\left.\left(6 F_1 F_3 \left(3 F_2{}^2+F_1 F_3\right) \dot{a} \left(9 \dot{F}_1 F_2{}^4+6 F_1 F_3 \dot{F}_1 F_2{}^2+F_1{}^2 \left(2 F_3 \left(F_3 \dot{F}_1+3 F_2 \left(\dot{F}_2-\bar{\rho} _d\right)\right)-3 F_2{}^2 \dot{F}_3\right)\right) F_2{}^2\right.\right.\right.\nn
\\
&\left.\left.\left.+a\left(-27 F_3 \dot{F}_1{}^2 F_2{}^8+18 F_1 F_3 \left(3 F_2{}^2 \ddot{F}_1-2 F_3 \dot{F}_1{}^2\right) F_2{}^6+18 F_1{}^2 F_3 \left(3 F_2{}^2 F_3 \ddot{F}_1-\dot{F}_1 \left(\dot{F}_3 F_2{}^2-2 F_3 \dot{F}_2 F_2\right.\right.\right.\right.\right.\right.\nn
\\
&\left.\left.\left.\left.\left.\left.+F_3{}^2 \dot{F}_1\right)\right) F_2{}^4+2 F_1{}^3 \left(9 \left(\dot{F}_3{}^2-F_3 \ddot{F}_3\right) F_2{}^4+18 F_3 \left(\bar{\rho} _d \dot{F}_3+F_3(t) \left(\ddot{F}_2-\dot{\bar{\rho}} _d\right)\right) F_2{}^3+6 F_3{}^2 \left(3 \bar{\rho} _d{}^2\right.\right.\right.\right.\right.\right.\nn
\\
&\left.\left.\left.\left.\left.\left.-3 \dot{F}_2 \bar{\rho} _d-3 \dot{F}_2{}^2-\dot{F}_1 \dot{F}_3+2 F_3 \ddot{F}_1\right) F_2{}^2+12 F_3{}^3 \dot{F}_1 \dot{F}_2 F_2-2 F_3{}^4 \dot{F}_1{}^2\right) F_2{}^2+F_1{}^4 F_3 \left(3 \left(\dot{F}_3{}^2-2 F_3 \ddot{F}_3\right) F_2{}^2\right.\right.\right.\right.\right.\nn
\\
&\left.\left.\left.\left.\left.+12 F_3\left(\left(\bar{\rho} _d+\dot{F}_2\right) \dot{F}_3+F_3 \left(\ddot{F}_2-\dot{\bar{\rho}} _d\right)\right) F_2{}^2+4 F_3{}^2 \left(3 \left(\bar{\rho} _d+\dot{F}_2\right) \left(\bar{\rho} _d-2 \dot{F}_2\right)-\dot{F}_1 \dot{F}_3+F_3 \ddot{F}_1\right) F_2+8 F_3{}^3 \dot{F}_1 \dot{F}_2\right) F_2\right.\right.\right.\right.\nn
\\
&\left.\left.\left.\left.-2 F_1{}^5 F_3{}^2 \left(F_2 \dot{F}_3-2 F_3 \dot{F}_2\right){}^2\right)\right)\right)/\left(a \left(3 F_2{}^2+F_1 F_3\right){}^2 \left(3 F_2{}^2+2 F_1 F_3\right)\right)\right)/\left(16 F_1{}^2 F_2{}^4 F_3{}^2\right)\,.
\end{align}

\end{document}